\newcommand{\CLASSINPUTbottomtextmargin}{0.9in}
\documentclass[journal]{IEEEtran}
\usepackage{amssymb}
\usepackage{amsmath}
\usepackage{amsthm}
\usepackage{epsfig}
\usepackage{graphicx}
\usepackage{wrapfig}
\usepackage{epstopdf}
\usepackage{algorithmic}
\usepackage{arevmath}     
\usepackage{balance}
\usepackage{caption}
\usepackage{tabularx}
\usepackage{color}
\usepackage{subcaption}
\usepackage{tikz}
\usepackage{multirow}
\usepackage{url}
\graphicspath{Figures/}
\usepackage[linesnumbered,ruled]{algorithm2e}
\SetKwComment{Comment}{/* }{ */}
\usepackage{array}
\theoremstyle{definition}

\newcolumntype{M}{>{\linespread{1}\selectfont\centering}m{2cm}}
\newcommand{\hide}[1]{}

\newcommand{\name}{\texttt{PePC}\space} 
\renewcommand{\name}{CPePC} 
\begin{document}

\title{\name: Cooperative and Predictive Popularity based Caching for Named Data Networks}
\vspace{-0.5cm}

\author{
    \IEEEauthorblockA{Pankaj Chaudhary*, Neminath Hubballi, Sameer G. Kulkarni} 
    
    \thanks{*Corresponding author: phd2001201004@iiti.ac.in \\
    Pankaj Chaudhary and Neminath Hubballi are with the Department of Computer Science and Engineering, Indian Institute of Technology Indore, India. \\
    Sameer Kulkarni is with the Department of Computer Science and Engineering, Indian Institute of Technology Gandhinagar, India.}
}

\maketitle

\begin{abstract}
Caching content is an inherent feature of Named Data Networks. Limited cache capacity of routers warrants that the choice of content being cached is judiciously done. Existing techniques resort to caching popular content to maximize utilization. However, these methods experience significant overhead for coordinating and estimating the popularity of content. To address this issue, in this paper, we present \name \space, which is a cooperative caching technique designed to improve performance. It accomplishes this through a combination of two factors. First, \name \space enhances efficiency by minimizing the overhead of popularity estimation. Second, it forecasts a parameter that governs caching decisions. Efficiency in popularity estimation is achieved by dividing the network into several non-overlapping communities using a community estimation algorithm and selecting a leader node to coordinate this on behalf of all the nodes in the community. \name \space bases its caching decisions by predicting a parameter whose value is estimated using current cache occupancy and the popularity of the content into account. We present algorithms for community detection, leader selection, content popularity estimation, and caching decisions made by the \name \space method. We evaluate and compare it with six other state-of-the-art caching techniques, with simulations performed using a discrete event simulator to show that it outperforms others.

\end{abstract}


\section{Introduction}\label{sec1}

The Internet traffic landscape has changed significantly in recent years, primarily due to the exponential growth in Internet users, an increase in the number of connected devices, the rise of over-the-top (OTT) streaming services (e.g., Netflix, YouTube TV), and the expansion of Internet of Things (IoT) applications \cite{Cisco2020}. The current TCP/IP Internet architecture, primarily designed for host-centric communication, is facing significant challenges in coping with the continuously rising volume of content-centric Internet traffic. This has sparked demand for the design of a new Internet architecture capable of effectively addressing the issues posed by the current host-centric approach. Many Information-Centric Networking (ICN) architectures \cite{Xylomenos2013} have been designed recently to support content-centric applications by caching the content in network routers to improve the user experience and reduce the load on the source. Named Data Networking (NDN) \cite{Zhang2014} is one of the most promising emerging ICN architectures, relying on name-based routing and caching operations. \\
\indent In the NDN, the content request and retrieval process are facilitated through two distinct packet types: \textit{Interest (Request)} and \textit{Data (Content)} packets \cite{Jacobson2009}. Consumers initiate the content retrieval process by sending an \textit{Interest} packet to the content provider (router/original source), which responds with a \textit{Data} packet containing the requested content.
NDN uses three primary data structures: \textit{Content Store (CS)} for caching content, \textit{Pending Interest Table (PIT)} tracks pending requests sent to upstream routers awaiting corresponding content, and \textit{Forwarding Information Base (FIB)} for storing routing information to forward requests to the next hop(s). \\
\indent As NDN supports network-level caching, allowing every network router to cache content \cite{Zhang2014}. Due to the limited storage capacity of routers, it is important to cache popular content \cite{Amadeo2021, Ashraf2024, Chaudhary2025, Hou2023} in the router that is often accessed by consumers in order to maximize content availability in the network. Content popularity estimation and caching decisions within the network can be made either cooperatively or non-cooperatively \cite{Zhang2015}. Routers regularly exchange data to identify popular content in a cooperative approach, but this introduces significant communication overhead in larger networks. On the other hand, in the non-cooperative approach, each router individually estimates popularity and makes caching decisions based on local information, which can result in content redundancy across the network and reduced hit rate. To this end, we propose a Cooperative and Predictive Popularity-based Caching (\texttt{\name}\footnote{A preliminary version of this paper appeared in IEEE CCNC 2024 \cite{Hubballi2024}}) to maximize the overall hit rate. We divide the network into several groups or communities and designate a leader node within each community to bring collaboration, which helps in popularity estimation, routing and caching decisions, and more. Our approach utilizes a predictive model inspired by Random Early Detection (RED) \cite{Floyd1993} to estimate the probability of caching content within the community. \texttt{\name} strategy caches content in the community based on the current router occupancy and the content's popularity. Community division in the proposed \texttt{\name} strategy helps cache content within each community by considering the interests of all consumers in that community, leading to more efficient content distribution. This community division mechanism reduces the burden on individual routers while also making efficient use of the network's available cache spaces. \\
The main contributions of this paper are as follows: \\
1. We propose \texttt{\name}, a novel cooperative predictive caching algorithm designed for content caching within the network.\\
2. We adopt a community detection algorithm to partition the network topology into multiple communities or clusters, and within each community, a leader node is designated to coordinate cooperative routing and caching. \\
3. We develop a parameter estimation technique which is inspired by Random Early Detection (RED) \cite{Floyd1993} to assist in caching decisions within each community. \\
4. We implemented \texttt{\name} on the popular Icarus~\cite{Saino2014} simulator. We compare the performance of the \texttt{\name} strategy with state-of-the-art caching techniques to show that it performs better. We also discussed the challenges associated with \texttt{\name} routing and caching algorithms for deploying in real-world NDN networks.

The structure of this paper is as follows: In Section \ref{sec2}, we present the analysis of traditional and recent advances in NDN caching techniques. Section \ref{sec3} first provides an overview of our proposed \texttt{\name} strategy, followed by detailed explanations of the community detection, cooperative routing, and the probabilistic caching method. Section \ref{sec4} outlines the results of the simulation studies. Finally, Section \ref{sec5} presents conclusions and outlines potential directions for future work.

\section{Literature Review} \label{sec2}
The support of in-network caching capabilities of the NDN architecture has drawn substantial research interest recently \cite{Khan2024, Li2024, Qu2023, Wang2024, Wu2023, Yoshida2024}. Although in NDN caching is an inherent feature, but there are other domains where such caching techniques are used. For example, many recent works \cite{Jin2024, Kim2025, Zhang2024} have used caching techniques in different domains to improve performance and efficiency by storing frequently accessed information. We conducted a considerable study on two distinct classes of NDN caching strategies: cooperative and non-cooperative strategies. These techniques primarily address cache lookup \cite{Zhang2015}, content placement \cite{Fan2020, Mahmood2018efficient}, content replacement \cite{Zhang2023}, and the static and dynamic nature of content popularity \cite{Naeem2020}.
\subsection{Cooperative Caching} In these caching methods, nodes in a network work together to optimize content dissemination and caching decisions. By leveraging collective knowledge, cooperative approaches aim to increase cache diversity,  improve cache hit, and reduce the load on the original content source. Nodes can achieve cooperation through either implicit or explicit means \cite{Zhang2015}. In the implicit technique, nodes do not explicitly advertise about the content to other nodes, whereas nodes using the explicit technique communicate actively with other nodes to share cache information. To explore the relevant works, we categorize cooperative decisions into two sub-classes: on-path and off-path approaches.
\subsubsection{On-path Caching}
Routers located on the transmission path between the content provider and the consumer make a collaborative decision (implicit or explicit) to perform the caching operations. On-path approaches improve content retrieval time and reduce network congestion by strategically placing content at different routers.
The works in \cite{Amadeo2022, Cho2012, Li2024, Yu2018} use on-path implicit cooperation strategies. \\
\textsl{WAVE~\cite{Cho2012}:} Introduces a popularity-based cooperative content placement strategy in which each downstream router along the content delivery path caches chunks based on the suggestion from the upstream router. \\
\textsl{Dynamic Popularity-Based Caching Permission (DPCP)~\cite{Yu2018}:} This strategy dynamically adjusts the threshold based on request patterns to prioritize caching popular content. To minimize redundancy, only popular content is cached on an alternate node along the download path when the cache is full. \\
\textsl{Max-Node Utility (MNU)~\cite{Khandaker2021}:} MNU is a producer-centric on-path caching strategy that selects the optimal router along the response path for content caching. This selection is based on the router's topological and dynamic characteristics. \\
\textsl{Popularity-Aware Closeness-based Caching (PaCC)~\cite{Amadeo2022}:} The PaCC strategy caches frequently requested content closer to consumers in the network. It uses an implicit cooperation mechanism by adding a field named PaC to the NDN Interest and Data packets to make caching decisions. Each router along the content delivery path compares its PaC with the PaC present in the Data packet to determine whether to cache the content. \\
\textsl{Caching-Resource Utilization-Based Strategy (CRUS)~\cite{Li2024}:} In CRUS, on-path routers in the downstream direction with the lowest resource utilization are selected to cache content. CRUS uses a probability threshold to differentiate between popular and non-popular content, prioritizing caching popular content closer to the consumer. Additional control fields are added to the Interest and Data packets to help select the on-path router with the lowest resource use.  \\
In \cite{Li2020}, a feedback-based cooperative content placement technique was proposed based on the relative popularity of the content from the downstream node. In \cite{Jaber2020}, the authors proposed a cooperative on-path caching technique that identifies the optimal router for content caching based on the distance from the source and the router's degree. Wu \emph{et al.} \cite{Wu2023} proposed an on-path caching technique in which routers in the downstream direction cache content by considering three parameters: content popularity, network topology, and content freshness. The authors in \cite{Kim2015} introduced a cache-capacity-aware mechanism for content-centric networks (CC-CCN) that performs selective caching and cache-aware routing. This strategy considers the cache capacity of on-path routers and the popularity of content to better utilize the in-network caches. CC-CCN exchanges information by modifying Interest and Data packets, reducing the communication overhead associated with packet exchanges. 
Recently, Wang \emph{et al.} \cite{Wang2024} presented a content caching technique for wireless sensor networks by leveraging the in-network caching feature of NDN. Sensor nodes estimate the caching weight based on multiple factors, such as content popularity, content lifetime, availability, and resource capability, to determine which content to cache.
\subsubsection{Off-path Caching} In this method, routers located outside the transmission path are involved in content searching and caching decisions. Routers can coordinate according to the predefined rules \cite{Saino2013b} or with the help of the Software-Defined Networking (SDN) controller \cite{Zhang2020} or the cluster head \cite{Huang2019}.\\
Saino \emph{et~al.} \cite{Saino2013b} proposed a hash-based routing and caching mechanism for retrieving and caching content at the targeted node. When an edge router receives a request from a consumer, it calculates the hash value of the requested content to locate the specified cache node in the network to redirect the request and cache the content in the return path. Another collaborative content caching scheme, CPHR \cite{Wang2015} is proposed by Wang \emph{et~al.} In this scheme, a hash function is used to partition the content space and subsequently map it to the appropriate caches of different nodes. This will improve the cache hit ratio and diversity in the network. The authors of \cite{Choi2014} designed a coordinated routing and caching scheme known as CoRC, which improves caching efficiency while reducing routing scalability issues by dividing the content namespace and assigning partitions to responsible routers using hashing techniques. 
In \cite{Mun2017}, the authors propose a collaborative caching method in which routers share cache summaries using a Bloom filter data structure within the neighborhood. This method helps utilize content available in off-path neighbors and reduces content redundancy in the neighborhood. Another work \cite{Chaudhary2025eencache} uses a Bloom filter to enhance content search and caching. To avoid false positives, it explores both on-path and off-path routers simultaneously. In some recent works \cite{Chaudhary2022, Chaudhary2023, Chaudhary2025, Reshadinezhad2023}, cooperative routing and caching methods have been developed to fetch content from off-path neighborhoods while forwarding the packet toward the original source. This is achieved through modifications to the Interest packet structure. Xu \emph{et~al.}~\cite{Xu2024} designed a content placement technique for satellite networks by dividing the network into multiple communities. This method identifies the optimal set of cache routers based on their connectivity within each community to reduce content transmission delays in satellite networks. Recent work \cite{Lee2025} designed a caching technique called Heartbeat for NDN to access off-path cached content. Instead of sending entire catalogs, it sends virtual interests to neighboring nodes whenever content is inserted or relocated in a cache. Heartbeat improves cache utilization while reducing bandwidth consumption and overhead.

\subsection{Non-cooperative Caching} 
Non-cooperative strategies are simplistic content caching strategies without having any extra coordination and communication overhead. In non-cooperative caching, each node between the content provider and the consumer makes individual decisions about caching and replacing the content. However, these approaches may result in content duplication in the network. \\
\textsl{Leave Copy Everywhere (LCE)~\cite{Jacobson2009}:} LCE is the most basic in-network caching technique for NDN architecture. LCE caches content in all cache-enabled routers between the content provider and requester without requiring coordination with other routers.  Due to the limited storage capacity of in-network routers, caching all content in the network at several locations leads to content redundancy, which may reduce the performance of the LCE. 
\\
\textsl{Prob(p)~\cite{Zhang2015}:} It is the simplest probability-based caching solution that overcomes the LCE strategy's content redundancy problem. In Prob(p), each router along the content downloading path randomly caches the content with a set probability value $p\in[0,1]$. When the value of $p$ reaches 1, the Prob(p) strategy becomes similar to LCE.

To summarize, both cooperative and non-cooperative caching techniques have several benefits and drawbacks depending on the application scenario. Cooperative caching improves cache hit ratios and reduces content redundancy while increasing signaling overhead and complexity of the algorithm, whereas non-cooperative caching is easier to implement but suffers from content redundancy, which reduces cache efficiency. Our preliminary work, PePC, considers cache availability and content popularity when making caching decisions. Preliminary results showed that caching decisions based on cache availability and popularity improve cache hit ratios and reduce content transmission time. However, the preliminary version is limited to individual routers making their independent decisions (routers do not share cache information with other routers), resulting in multiple copies of the same content being cached in the network. Additionally, traditional routing and caching techniques in NDN lead to inefficient utilization of caching resources and increase the load on the server. Therefore, in this article, we extend our preliminary work to integrate cooperation for routing, caching, and popularity estimation, aiming to better utilize the limited capacity of routers. To bring cooperation amongst the routers, \texttt{\name} divides the network into several communities. Further, to reduce the signaling overhead, \texttt{\name} adds additional fields to the NDN packets. These fields are used for efficient content searching and caching within the community. Routers coordinate with the community leader using the changed packet structure to make informed caching decisions. These additional fields in NDN packets can help to reduce the number of messages that need to be exchanged within the communities for content searching and caching.

\section{Proposed Caching Technique} \label{sec3}

\begin{figure*}[bthp]
 \centering
  \includegraphics[width= 0.7\textwidth]{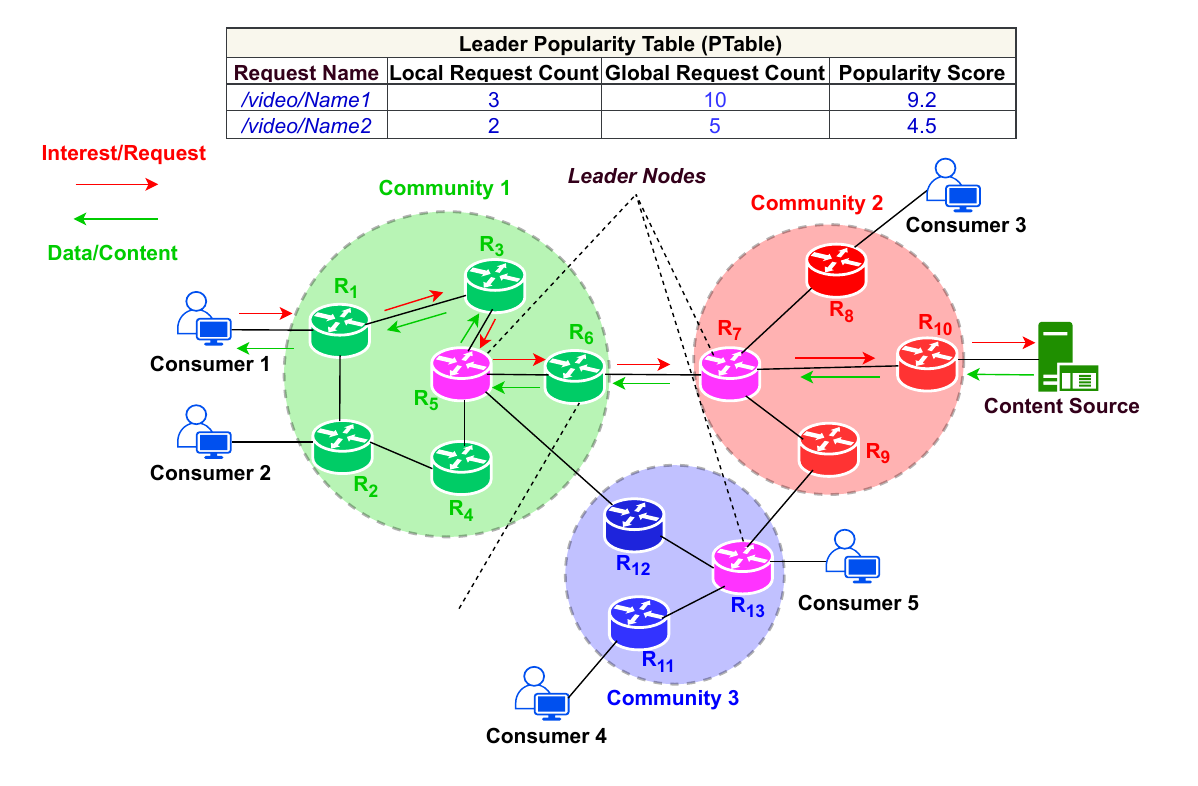}
  \caption{\texttt{\name} Reference Architecture}
   \label{fig:proposedarchitecture}
\end{figure*}

We begin this section with an overview of our proposed cooperative \texttt{\name} caching technique. Following that, we introduce detailed design and relevant algorithms for community formation, local and global popularity estimation, request routing, and the design of predictive caching decisions for content placement. We explain the functioning of these algorithms with the help of Fig.~\ref{fig:proposedarchitecture}. Table \ref{tab: notations} presents an overview of notations that we use while describing the \texttt{\name} technique.

\begin{table}[bthp]
 \centering
 \caption{Notations used in the Description of \texttt{\name}}
 \label{tab: notations} 
\scalebox{0.8}{
    \begin{tabular}{|c|c|}
     \hline   
    \textbf{Notations} & \textbf{Description} \\
    \hline
    $G$ & Network graph \\
    \hline
    $V$ & Set of nodes in the network \\
    \hline
    $R$ & Set of cache-enabled routers \\
    \hline
    $C$ & Set of communities\\
    \hline
     $N$ & Number of communities \\
     \hline
    $L$ & Set of leader nodes \\
    \hline
    $ \lambda, \omega \in [0,1]$ & Weight parameters \\
    \hline
    $P(Name)$ & Popularity of content \texttt{Name} \\
    \hline
    $L_f$, $G_f$ & Local and Global request counters\\
    \hline
    $PTable$ & Popularity table of $L_i$ \\
    \hline
    M & Number of distinct contents in $PTable$ \\
    \hline
    T & Period for popularity estimation \\
    \hline
    $\mathcal{A}_{R_i}$ & Average cache occupancy of $R_i$ \\
    \hline
    $\Theta_{R_i}$ & Current cache occupancy of $R_i$ \\
    \hline
    $\mathcal{RO}$ & Relative cache occupancy \\
    \hline
    $min_{th}, max_{th}$ & Minimum and maximum cache thresholds \\
    \hline
    $\rho_1 , \rho_2 \in [0,1]$ & Tunable parameters for cache thresholds \\
    \hline
    $S_{R_i}$ & Total storage capacity of the router $R_i$ \\
    \hline
    $P_1, P_2$ & Caching decision values \\
    \hline
    $P_{max}$ & Maximum value for $P_1$ \\
    \hline
    $\beta$ & Counter between cached and non-cached units\\
    \hline
     $\Delta$ & Popularity threshold \\
     \hline
    \end{tabular} } 
  \vspace{-4mm}
\end{table}

\subsection{Design Rationale} The content caching feature of NDN/ICN architecture improved content availability and response time by caching the content in network routers. However, as the router's capacity is very limited, caching all content at all times does not optimize the performance of caching methods. As a result, the decision to cache the content in the router, depending on the consumer request pattern, becomes critical for any caching method. Cooperation among routers optimizes content delivery and caching decisions. However, in order to make a cooperative decision, each router must actively exchange its cache state and content popularity information with the other nodes, which may result in communication overhead. On the other hand, non-cooperative strategies emphasize individual node-centric decisions, which reduces the communication overhead problem. However, the main issue associated with this technique is that routers located on the request or response path are unaware of the availability of content outside the path, which diminishes the performance of caching techniques. To this end, we design \texttt{CPePC} caching technique that encourages cooperation among the nodes by dividing the network topology into several communities. A leader node is elected in every community region to achieve intra-domain and inter-domain communication. Routing and caching through the leader node enhance cache utilization by making use of content available at both on-path and off-path routers. Additionally, exchanging popularity information via the leader node within and among other communities significantly reduces communication overhead. Further, \texttt{CPePC} is reactive to cache occupancy, where it caches popular content as occupancy increases. The motivation behind dividing the network into multiple communities and measuring the popularity of content is to ensure that users in the same region have quick access to locally trending content, for example, local news or popular videos.

\subsection{\texttt{CPePC} Packet Structure} 
\begin{figure}[bthp]
 \centering
  \includegraphics[width= 1.0\columnwidth]{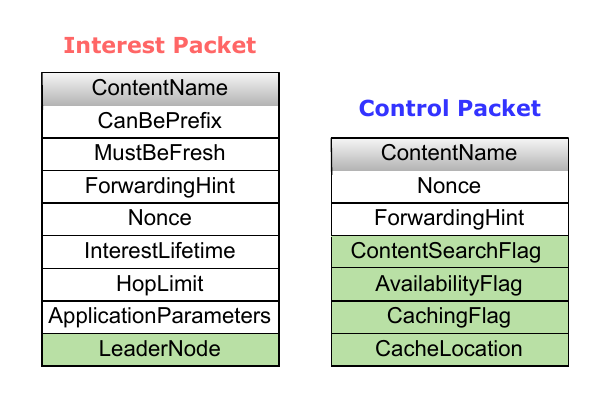}
  \caption{\texttt{CPePC} Interest Packet Structure (IPS) and Control Packet Structure (CPS).}
  \label{fig:cpepcPacketStructure}
  \vspace{-4mm}
  \end{figure}

\texttt{CPePC} uses separate packets for communicating with the community leader and for forwarding requests towards the content provider using next-hop information. The Interest Packet Structure (IPS) is used to forward requests from one router to another router, while the Control Packet Structure (CPS) facilitates communication between a router and the community leader. We added a field to the native NDN Interest packet (IPS) format to maintain leader information. Additionally, we designed a new packet structure, CPS, to optimize content searching and caching within the community, while aligning with the standard NDN packet format. The fields marked in green, as shown in Fig.~\ref{fig:cpepcPacketStructure}, are newly added to the native NDN packet format specification (version 0.3) \cite{ndninterestpacket}. In CPS, the \textit{Nonce} and \textit{ForwardingHint} fields serve the same purpose as in the native NDN Interest packet format. Below, we elaborate on the details of these extended fields.

\begin{itemize}
    \item \textit{ContentName:} Specifies the name of the content being requested or cached.
    \item \textit{Nonce:} Uniquely identifies a packet.
    \item \textit{LeaderNode:} Contains the information of the leader node.
     \item \textit{ForwardingHint:} Provides the forwarding path for content retrieval within the community.
    \item \textit{ContentSearchFlag:} This field indicates whether the content should be searched within the community with the help of the leader. If the value of this field is set to \texttt{True}, the request is forwarded to the leader.
    \item \textit{AvailabilityFlag:} Indicates whether the requested content is available within the community. If this flag is \texttt{True}, the leader responds to the router by setting the \textit{ForwardingHint} field.
    \item \textit{CachingFlag:} This field is set to \texttt{True} when the router receives a Data packet and the content is not available in its cache.
    \item \textit{CacheLocation:} Specifies the location within the community where the content is to be cached by the leader.
\end{itemize}

\subsection{Community Formation}
In order to establish cooperation among the routers, we use a graph representation to divide large-scale network topology into several communities or clusters\footnote{We use 'community' and 'cluster' interchangeably throughout the paper.}. In a network graph G, nodes typically represent source nodes (the original storage of contents), a set of consumer nodes (making content requests), and a set of routers capable of caching content, while the edges define the connections or links between these nodes. 

To generate a group or community, we use the Louvain algorithm \cite{Blondel2008}, a well-known community detection approach, to partition a network topology into multiple communities or clusters.  An example network topology is shown in  Fig \ref{fig:proposedarchitecture}. where three such node communities are indicated. Louvain algorithm is well-known for effectively dividing large networks and finding non-overlapping communities. \\
This algorithm operates through two iterative phases, with the goal of improving the network's community structure by optimizing the modularity\footnote{It is used to assess the quality of the community structure within a network.} score. The first phase of the algorithm is known as modularity optimization; in this phase, the algorithm begins with each node forming its own community in the network graph G. So, the number of communities in G equals the number of nodes in G. Then, it computes the modularity score and iteratively checks each node by calculating the gain in modularity score from joining the node from nearby community $C_i$ to $C_j$. Nodes from community $C_i$ will only join community $C_j$ if it results in an increase in the modularity score. This phase terminates when no further gain in modularity can be achieved by moving the node from one to another. The algorithm's second phase is the community aggregation phase; during this phase, the communities generated in the first phase are considered nodes in a new network. Again, modularity optimization is performed on this new community network, and if there is a gain in modularity, the node is moved from one community to another. The process iterates between these two phases until no further improvement in modularity is achievable by combining communities. The algorithm terminates by creating the communities, and each node is assigned to one of the communities. \\

\noindent\textbf{Leader Node Selection:} Following the division of the network graph G into N communities, we select a leader node for each community to facilitate cooperation among the communities to ensure efficient handling of requests and contents. This node will be in charge of exchanging the details of cached content and the popularity of the content and also dictate the caching decisions for all the nodes within the community. We select the node with the highest betweenness centrality as the leader node within each community. The reason for choosing the node with the highest betweenness centrality as the community’s leader is twofold: first, because of its higher centrality, the leader can efficiently communicate with other nodes (less number of messages) in the community and also acts as a bridge between different communities of the network, effectively reducing communication overhead; second, if some node disconnects from the leader due to link failure, the leader can still establish communication with the affected node via an alternate path. If multiple nodes have the same number of betweenness centrality, the node with the minimum average distance from each node within the community is selected as the leader of that community. If the leader node of any community fails, the next node within the community with the highest betweenness centrality is elected as the new leader.
The leader node $L_i$ of any community $C_i$, $i \in \{1, 2, \ldots, N\}$, is selected as given in Equation \ref{leaderselection}.

\begin{equation}
\label{leaderselection}
    L_i = \underset{v \in C_i}{\arg\max} \{BC(v)\}
\end{equation}
where $L_i$ represents the leader node of community $C_i$, and $BC(v)$ signifies the betweenness centrality of node $v$ within community $C_i$. 

\begin{figure*}[!bthp]
 \centering
  \includegraphics[width= 0.7\textwidth]{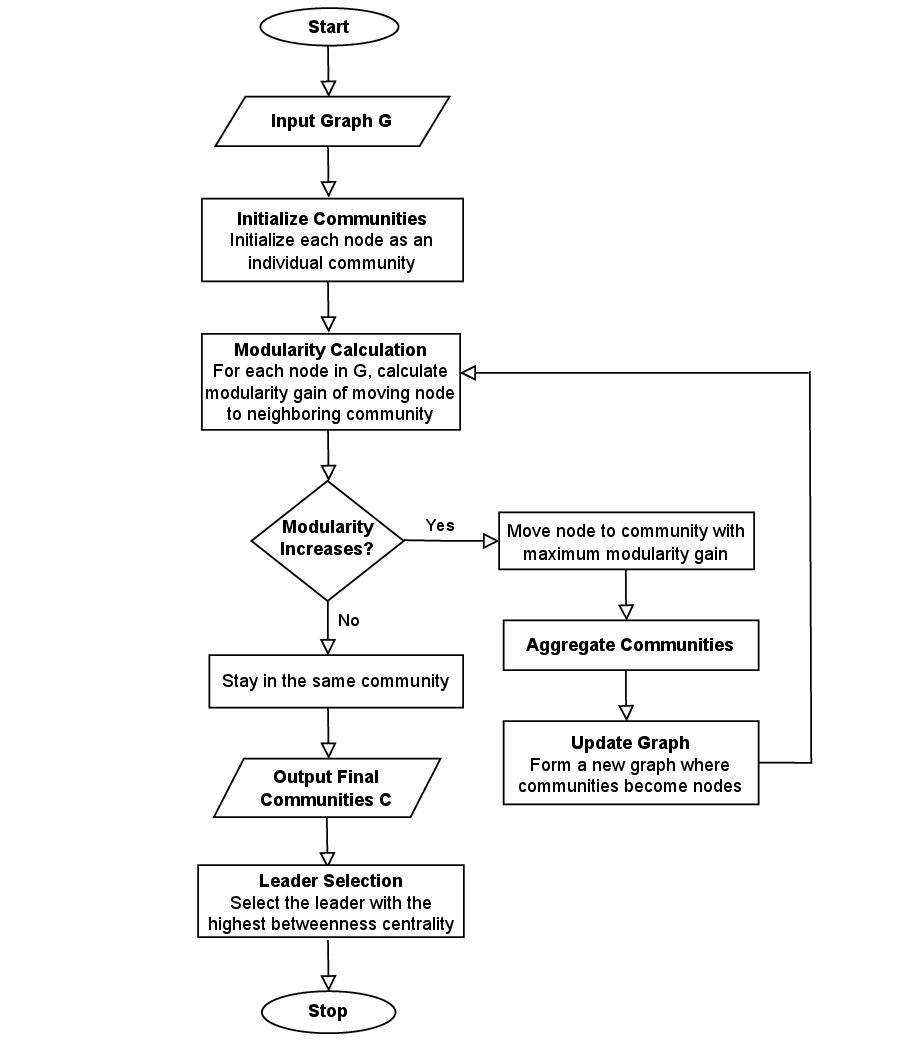}
  \caption{Community Formation and Leader Selection Process}
   \label{fig:communityformation}
   \vspace{-4mm}
\end{figure*}

Fig.~\ref{fig:communityformation} illustrates the process of community formation of the graph network topology $G$ and the selection of a leader node $L_i$ within each community.

\subsection{Popularity Estimation} Content popularity estimation plays an important role in NDN/ICN networks to improve the performance of the system, as the content would be cached in the routers. The popularity of content is determined by the number of times the routers receive content requests. Based on consumer usage patterns, the content will be classified as either locally popular or globally popular. If the content is popular in a particular geographical region or community, it is termed as locally popular, and if it is requested by a wide range of consumers from many geographic locations, it is considered globally popular. In order to optimize the limited router storage, it is important to consider both local and global popular content for caching decisions. Our \texttt{\name} method considers both local and global request patterns while making a caching decision.  To achieve this, the leader node of the community maintains a popularity table (PTable), which keeps track of all the content requests inside or outside the community. By maintaining this table only at the leader node, we reduce the burden on every other router within the community. It is worth noting that the popularity of content fluctuates over time within specific regions or communities, influenced by consumer interests. An efficient caching approach should periodically update content popularity over a fixed time \texttt{T} to accommodate these changes effectively. To measure popularity over time, we employ the Exponential Weighted Moving Average (EWMA) \cite{Gui2020, Lucas1990} model, which considers both present usage patterns and historical trends. Consumer content access patterns remain relatively stable over short time intervals, so frequently updating the popularity does not provide benefit because it increases the communication overhead due to information exchange and also content replacement rates.  To facilitate the popularity estimation of content, the leader of each community maintains two variables, $L_f$ and $G_f$ which track the local and global frequency of requests for content \texttt{Name}. $L_{f_i}$ indicates the frequency of \texttt{Name} within a community $C_i$, and $L_{f_j}$ indicates the frequency of \texttt{Name} from community $C_j$ where $C_i \neq C_j$.
$L_{f_i} \forall i\in C$ collectively represent the global popularity $G_f$ of the requested content \texttt{Name}.  Algorithm~\ref{popularityupdate} outlines the approach the leader node $L_i$ uses within each community $C_i$ to monitor the request count $L_f$ in the PTable. When a new content type is requested, an entry is created in the PTable for that content (lines 9 to 11), and if an entry already exists in the PTable, the counter value is incremented by one (lines 5 to 7). Each leader node exchanges its PTable with other leader nodes at intervals of \texttt{T} to update the $G_f$. PTable stores four types of information: \texttt{Name} (content request name), $L_f(Name)$ (count of content requests within community $C_i$), $G_f(Name)$ (count of content requests across communities), and $P(Name)$ (popularity score of the content, calculated using Equation~\ref{popularityestimation}). The structure of PTable is shown in Figure~\ref{fig:proposedarchitecture}. The EMWA approach is used in Equation \ref{popularityestimation} to calculate the popularity of the content $P(Name)$.
\begin{equation}   
\label{popularityestimation}
    P_T(Name) = \lambda \times P_{T-1}(Name) + (1-\lambda) \times G_f(Name)
\end{equation}
where  $P_T(Name)$ denotes the popularity of content \texttt{Name} over the period \texttt{T}, $P_{T-1}(Name)$ represents the estimated popularity of content \texttt{Name} recorded in the PTable during the previous time step \texttt{T-1}, $G_f$ is the count of the number of requests for the content \texttt{Name} over the period \texttt{T}, and $\lambda \in [0, 1]$ is the EMWA smoothing factor. 
In order to adapt to the changing patterns of consumer requests over time, we compute a dynamic popularity threshold $\Delta$ within each community instead of the static threshold to identify the content as popular or unpopular. Once the content has reached the threshold $\Delta$, it is considered popular. The dynamic popularity threshold $\Delta$ is computed over the time period \texttt{T} shown in Equation \ref{popularitythreshold}.

\begin{equation}
\label{popularitythreshold}
    \Delta = \frac{ \sum\limits_{k=1}^{M} {P(Name_k)}} {M}
\end{equation}
where $\Delta$ represents the dynamic popularity threshold of community $C_i$, $P(Name_k)$ represents the popularity of content \texttt{Name} during time \texttt{T}, and M represents the total number of distinct contents in the PTable during that time interval. 

\begin{algorithm}[bthp]
\small
\caption{Estimating Request Count at $L_i$}
\label{popularityupdate}
  \begin{algorithmic}[1]
    \STATE $L_f \leftarrow 0$ // Local Request Counter
      \STATE PTableOf($L_i) \leftarrow \{\}$
     \STATE \texttt{Name} $\leftarrow$ Content Request in $C_i$ 
     \FOR{each \texttt{Name} in Community $C_i$}
      \IF{\texttt{Name} $\in$ PTableOf($L_i$)}
       \STATE $L_f(\texttt{Name})++$
       \STATE Update PTableOf($L_i$)
        \ELSE
         \STATE Create Entry for \texttt{Name} in PTableOf($L_i$)
         \STATE $L_f(\texttt{Name}) = 1 $
         \STATE Update PTableOf($L_i$)
        \ENDIF
        \ENDFOR      
  \end{algorithmic}
\end{algorithm}
The popularity table so computed by the Algorithm \ref{popularityupdate} is exchanged by leader nodes with other leaders every \texttt{T} time period as shown in Algorithm \ref{popularitycluster}.

\begin{algorithm}[bthp]
\small
\caption{PopularityOf($C_i,Name$)}
\label{popularitycluster}
\begin{algorithmic}[1]
\STATE $L_{f_i}(\texttt{Name}) \leftarrow$ Request Count of $C_i$
\STATE $L_{f_j}(\texttt{Name}) \leftarrow$ Request Count of $C_j$
\FOR{each $C_i \in C$}
  \FOR{each Time interval \texttt{T}}
   \STATE Receive PTableOf($L_j$) from $C_j$
    \STATE $G_f(\texttt{Name}) \leftarrow L_{f_i}(\texttt{Name}) + L_{f_j}(\texttt{Name})$
    \STATE Update PTableOf($L_i$)
    \STATE $P(Name) \leftarrow $ Update Popularity as in Equation \ref{popularityestimation}
    \ENDFOR
    \ENDFOR \\
    \STATE Return $P(Name)$
\end{algorithmic} 
\end{algorithm}

\subsection{Request Routing} The content search within any community $C_i$ is performed through the leader of the respective community. The content request handling procedure of the \texttt{CPePC} is shown in Algorithm \ref{routingrequest}. 
Whenever a router $R_i$ receives a request for content \texttt{Name}, it does the following: first, it checks its cache; if the content is available in its cache, it responds with the content. Otherwise, $R_i$ forwards the request to either the leader node of the community or the next hop router along the shortest path to the content source. The forwarding process within each community can be controlled by including the leader information in the \textit{LeaderNode} field of the IPS and setting the \textit{ContentSearchFlag} field to \texttt{True} or \texttt{False} in the CPS, as shown in Fig.~\ref{fig:cpepcPacketStructure}. Note that the fields added to the CPS are enabled only when communication between the router and the community leader takes place, while packet forwarding to the upstream router towards the content provider is accomplished through IPS.

Each router within the community knows its respective leader node. So, when router $R_i$ receives the request, it checks the \textit{LeaderNode} field of IPS. If the leader node specified in the \textit{LeaderNode} field matches the leader of $R_i$, $R_i$ can directly forward the request to the next upstream router (line 13 in Algorithm \ref{routingrequest}), as it means that the leader node has already been visited by the previous hop router within the same community as $R_i$. Otherwise, it forwards the request to the leader node by setting \textit{ContentSearchFlag} to \texttt{True} and searches for the content within the community with the help of the leader (lines 4 to 10 in Algorithm \ref{routingrequest}). Upon receiving the request, the leader node responds with one of three types of messages: firstly, if the content is available in the leader's CS, it replies with content (Data packet); secondly, if the content is available in router $R_j$ (Router $R_j$ is in the same community as $R_i$) within community $C_i$, it replies with the location of $R_j$ (set the \textit{ForwardingHint} field in CPS to specify the path an Interest packet follows to retrieve the Data); and thirdly, if the content is not available within the community, it responds with message NDATA (\textit{AvailabilityFlag=False}) which Indicates that the requested content is unavailable within the specific community. When $R_i$ receives the second type of message, it routes the request to $R_j$ using the path specified by \textit{ForwardingHint} to fetch the content. When $R_i$ receives the third type of message, it appends the visited leader node information to the \textit{LeaderNode} field and routes the request to the next hop router by consulting the FIB.

\begin{algorithm}[bthp]
\small
\caption{Handling Content Request at $R_i$}
\label{routingrequest}
  \begin{algorithmic}[1]
     \STATE \texttt{Name} $\leftarrow$ New Content Request at $R_i \in C_i$     
     \IF{\texttt{Name} $\in$ CacheOf($R_i$)}
        \STATE Serve from $R_i$
     \ELSIF{Leader $L_i$ of $R_i$ $\notin$ \textsl{LeaderNode}}
        \STATE Forward Content Request to $L_i$ $\in$ $C_i$
         \IF{\texttt{Name} $\in$ CacheOf($L_i$)}
            \STATE Serve from $L_i$
        \ELSIF{\texttt{Name} $\in$ CacheOf($R_j \in C_i$) with $R_i \neq R_j$}
              \STATE Serve from $R_j$
        \ENDIF
     \ELSE
     \STATE \textsl{LeaderNode}.append($L_i$)
     \STATE Forward Content Request to Next Hop Router\\ on the Shortest Path to Source
     \ENDIF
  \end{algorithmic}
\end{algorithm}

This procedure is performed in each community until the content is found. In \texttt{CPePC}, only the first entry router of each community is responsible for coordinating with the leader node during request forwarding, due to the \textit{LeaderNode} field of IPS. This approach reduces communication overhead, as routers in the path do not need to coordinate with the leader. With the extension of the NDN packet structure, \texttt{CPePC} enables efficient content retrieval both within and outside the community. For example, consider the network topology shown in Fig.~\ref{fig:proposedarchitecture}. When router $R_1$ receives a content request from \texttt{Consumer1}, it forwards this request to the leader node, which in this case is $R_5$. Assuming that the content \texttt{Name} is available in \texttt{Community1} at $R_4$, after receiving the reply from leader $R_5$, $R_1$ forwards it to $R_4$ via this path: $R_1 \rightarrow R_2 \rightarrow R_4$. 

\subsection{Caching the Content} Here, we discuss how \texttt{\name} method decides to cache the content. This is a fundamental aspect of any caching scheme, as it determines how efficiently the limited cache space can be utilized and managed to maximize content availability in the network. This includes two major components: caching and replacement strategies. Caching decisions involve deciding which content to cache and where to cache it, while replacement strategies indicate which content to evict to make room for the new arrival. In order to avoid content redundancy within the community, our proposed \texttt{\name} technique performs both content caching and content replacement operations in a collaborative manner with the support of the leader node of the community. 

\noindent \textbf{(i) Caching Method:}
The content caching decision of our proposed \texttt{\name} strategy draws inspiration from the Random Early Detection (RED) \cite{Floyd1993} queue management technique. In order to reduce network congestion, RED uses a probabilistic decision-making approach to mark or unmark packets as they arrive at a queue based on queue occupancy by setting predefined values of minimum and maximum thresholds. We design a cooperative predictive decision-making algorithm to cache content in the community. Our primary objective in designing such a caching scheme is to enhance the network's cache hit ratio by efficiently caching diverse content within the community. The proposed \texttt{\name} technique allows the router to cache the content based on the average cache occupancy. The average cache occupancy of the router fluctuates dynamically, which benefits in deciding whether the router can cache the content based on its current occupancy. The average cache occupancy of the router is updated using Equation \ref{cacheoccupancy}.
\begin{equation}
\label{cacheoccupancy}
    \mathcal{A}_{R_i} = (1- \omega) \cdot \mathcal{A}_{R_i, old} + \omega \cdot \Theta_{R_i}
\end{equation}
where $\mathcal{A}_{R_i}$ represents the current average cache occupancy of $R_i$, ${A}_{R_i, old}$ represents the last updated cache occupancy value, $\omega$ represents the cache weight factor, and $\Theta_{R_i}$ represents the current cache occupancy of $R_i$. We employ minimum ($min_{th}$) and maximum ($max_{th}$) thresholds for comparison with average cache occupancy in the caching decision process. These thresholds are determined based on the total capacity of the router, as illustrated in Equations \ref{minth} and \ref{maxth}. The $min_{th}$ and $max_{th}$ represent the reserved fractions of space allotted for content caching, reflecting the lower and upper bounds.
\begin{equation}
    min_{th} = \rho_1 \cdot S_{R_i} 
    \label{minth}
\end{equation}
\begin{equation}
    max_{th} = \rho_2 \cdot S_{R_i}
    \label{maxth}
\end{equation}
Here, $S_{R_i}$ represents the total storage capacity of the router $R_i$, while $\rho_1$ and $\rho_2$ are tunable parameters that range between 0 and 1 to adjust the minimum and maximum cache thresholds. These parameters are adapted based on content caching decision-making and the available storage capacity of the router.

Similar to request routing, we have also optimized the caching process through communication with the community leader. In the downstream direction, when the content arrives at router $R_i$ within community $C_i$, $R_i$ forwards one copy toward the consumer to avoid delay caused by caching decisions, and another copy to its leader to potentially cache the content by setting the \textit{CachingFlag} to \texttt{True} in the CPS, as shown in Fig.~\ref{fig:cpepcPacketStructure}. Note that if the on-path router $R_i$ already has the content, it will not forward a copy of the content to the leader. Upon receiving the request, the leader node decides whether to cache the content. If the content is to be cached, it replies by setting the \textit{CacheLocation} field to the selected router. This decision assumes that the leader is aware of the locations and occupancy levels of routers within the community through coordination. Algorithm \ref{contentcaching} describes the content caching decision made by the leader node upon receiving a new content \texttt{Name}. When content is unavailable within the community, the leader node caches the content at a selected router within the community when any one of the three conditions mentioned below is met; otherwise, the content is discarded.
Note that the content will be cached at one of the on-path routers within each community during content transfer between the provider and the consumer if it is not cached anywhere in the community. To reduce latency in content delivery, \texttt{CPePC} adopts a simple on-path caching technique that caches content only at the router located on the content transmission path. This content placement scheme pushes content closer to the requested consumers without adding the computational overhead of selecting a router within the community for caching. \\
\noindent\textsl{(i) Average cache occupancy is lower than the minimum threshold:} According to the condition shown in Equation \ref{case1}, the leader node $L_i$ caches all incoming contents in selected router $R_i \in C_i$  regardless of their popularity in order to make the best use of the available cache space. As cache occupancy increases, the frequency of caching popular content increases significantly, while unpopular content is evicted from the cache.
\begin{equation}
\label{case1}
    \mathcal{A}_{R_i} < min_{th}
\end{equation}

\noindent\textsl{(ii) Average cache occupancy is between the minimum and the maximum threshold:} When the cache occupancy lies between the minimum and maximum thresholds, as indicated in Equation \ref{case2.1}, the leader $L_i$ only caches the content at the selected router $R_i \in C_i$ when condition $\mathcal{RO} \geq P_2$ is satisfied; otherwise, the content is discarded. 
\begin{equation}
\label{case2.1}
     min_{th} \leq \mathcal{A}_{R_i} < max_{th}
\end{equation}

\begin{equation}
 \label{case2.2}
     P_1 = \frac{P_{max}(\mathcal{A}_{R_i} - min_{th})}{(max_{th} - min_{th})}   
\end{equation}

\begin{equation}
 \label{case2.3}
     P_2 = \frac{P_1}{1 - (\beta \cdot P_1)}
\end{equation}
Equations \ref{case2.2} and \ref{case2.3} are used to calculate the early prediction parameters $P_1$ and $P_2$ for the content caching decisions. The value of $P_1$ ranges between 0 and $P_{max}$, where $P_{max}\in[0, 1]$ represents the upper limit for $P_1$, and $\beta $ denotes the counter between cached and uncached content units. When content is not cached, the value of $\beta $ is reset to zero.
\begin{equation}
     \mathcal{RO} = \frac{P(Name, L_i)}{max(P(Name_k, L_i))}
    \label{case2.4}
\end{equation}
$\mathcal{RO}$ is a numerical value ranging from 0 to 1, which signifies the relative cache occupancy of content within community $C_i$, computed as shown in Equation \ref{case2.4}. In the PTable of leader $L_i$ within community $C_i$, $P(Name, L_i)$ represents the popularity of content \texttt{Name}, while $max(P(Name_k, L_i))$ denotes the maximum popularity value among all contents in the PTable.

\noindent\textsl{(iii) Average cache occupancy reaches the maximum threshold:} When the average cache occupancy reaches the maximum threshold value, as shown in Equation \ref{case3}, it becomes necessary to store only the most frequently accessed content by consumers in order to improve content availability in the community. When the value of $P(Name)$ is greater than that of $\Delta$, leader node $L_i$ caches the content at a specific router $R_i$.

\begin{equation}
   \mathcal{A}_{R_i} \geq max_{th}
    \label{case3}
\end{equation}

\begin{algorithm}[bthp]
\small
\caption{Caching Content within Community $C_i$}
\label{contentcaching}
\begin{algorithmic}[1]
    \STATE \texttt{Name} $\leftarrow$ New Content at $R_i$
     \STATE Size $\leftarrow$ SizeOf( \texttt{Name})
       \STATE $\Theta_{R_i}$ $\leftarrow$ CurrentCacheOccupancyOf($R_i$)
       \IF{\texttt{Name} $\notin$ CacheOf($R_i$) }
       \STATE Forward Content \texttt{Name} to $L_i$ $\in$ $C_i$
      \IF{\texttt{Name} $\notin C_i$}
     \IF{$\mathcal{A}_{R_i}$ $<$  $min_{th}$}
        \STATE Cache \texttt{Name} at $R_i$ //maximize cache utilization.
        \STATE Update $\Theta_{R_i}$
     \ELSIF{$min_{th}$ $\le$ $\mathcal{A}_{R_i}$ $<$ $max_{th}$}
     \STATE Cache \texttt{Name} at $R_i$ if $\mathcal{RO} \ge {P}_2$
     \STATE Update $\Theta_{R_i}$
     \ELSE
     \STATE $P(Name)$ $\leftarrow$ PopularityOf($C_i, Name$)
      \STATE Cache \texttt{Name} at $R_i$ if $P(Name) \ge \Delta$
       \STATE Update $\Theta_{R_i}$
   \ENDIF
    \ENDIF
   \ENDIF

\end{algorithmic} 
\end{algorithm}

\noindent \textbf{(ii) Replacement Strategy:} When the leader node chooses a specific router within the community to cache the content, but the capacity of the selected router is full, the \texttt{\name} invokes the replacement algorithm to evict the less frequently accessed content from the CS and cache the fresh popular content. LRU performs well in scenarios where content access patterns exhibit high temporal locality. However, its performance degrades in cases of poor temporal locality in access patterns, such as requests made by consumers in a round-robin fashion \cite{Chaudhary2023}. Therefore, we also explored various replacement algorithms, such as Perfect Least Frequently Used (PLFU) and Random policy, to address these challenges. PLFU tracks access frequencies for all content requests, even if the content is not currently in the cache, and evicts the content with the lowest access frequency, while Random policy chooses the content to evict at random without regard for access frequency. PLFU requires an additional data structure to maintain the access frequency of the content, making it computationally more expensive compared to LRU and Random. Additionally, search and replacement tasks cannot be performed in constant time, as the router needs to compare the frequency of the newly arrived content with the content already in the cache.

\section{Experiments and Evaluation} \label{sec4}

In this section, we provide details about the simulation setup, performance metrics, and network topologies used for performance evaluation,  followed by the simulation results. We compared our proposed \texttt{\name} strategy with five well-known state-of-the-art caching techniques: LCE~\cite{Jacobson2009}, DPCP~\cite{Yu2018}, MNU~\cite{Khandaker2021}, PaCC~\cite{Amadeo2022}, CRUS~\cite{Li2024} as well as with the results of the preliminary version PePC~\cite{Hubballi2024} of the \texttt{\name} method.

\subsection{Simulation Setup} In order to assess the performance of our proposed \texttt{\name} caching method, we use the Icarus \cite{Saino2014} simulator. Icarus is a Python-based discrete-time event simulator explicitly designed for evaluating ICN/NDN routing and caching techniques. It offers a packet-level simulation environment, enabling users to analyze the behavior of individual packets within a network. In our simulations, content requests are generated using a Poisson distribution with 10 requests per second, which means that 10 requests are expected every second on average. Similar to previous studies \cite{Amadeo2021, Man2021, Wang2024a}, we adopted the Zipf distribution \cite{Breslau1999} to model the popularity of content by varying the skewness parameter $\alpha$, which ranges from 0.6 to 1.2. The web user request pattern is closely related to Zipf popularity distribution \cite{Breslau1999}, allowing caching techniques to determine appropriate cache sizes to achieve the desired cache hit ratio and diversity. \\ \indent The results were generated using the following simulator settings unless stated otherwise: the simulation content catalog contained $10^4$ distinct contents, uniformly distributed among sources. The initial phase involved a warm-up period \cite{Gui2020} of $5 \times 10^4$ requests to reach a stable state, which were not included in the performance evaluation. Following the warm-up phase, all consumers in the network requested $10^5$ contents, and these requests were considered for performance evaluation. Assuming that each router in the topology is capable of caching content with the same capacity. The storage capacity of routers varies between 0.05\% to 0.25\% of the simulation catalog size. Routers use the LRU algorithm to evict content and make room for new arrivals. In the network, all packets are forwarded toward the upstream and downstream nodes using Dijkstra’s weighted shortest path algorithm. All simulations are conducted on large-scale RocketFuel \cite{Spring2002} ISP topologies, and the details of these topologies are provided in the following subsection. The simulation parameters for both the \texttt{\name} and PePC methods are configured as follows: the minimum and maximum threshold parameters ($\rho_1$ and $\rho_2$) are set to 0.2 and 0.6, respectively; $P_{max}$ set to 1; $\lambda$ is set to 0.125 to prioritize recent data over historical data; weight factor $\omega$ for cache occupancy estimation is set to 0.125, and \texttt{T} is set to 10 seconds. Every simulation is run ten times, and the plots are generated using 95\% confidence intervals. Table \ref{tab: experimental} summarizes the most important standard simulation parameters.

\begin{table}[htbp]
\centering
 \caption{Simulation Settings}
 \label{tab: experimental} 
 \scalebox{0.8}{
    \begin{tabular}{|c|c|}
     \hline   
      \textbf{Parameters} & \textbf{Value} \\
      \hline
       Network topology & Exodus (AS 3967) and Abovenet (AS 6461) \\
      \hline
       Content catalog size & $10^{4}$ objects \\
      \hline
       Number of requests & $10^{5}$ objects  \\
      \hline
       Number of warm-up requests & $5 \times 10^4$ objects  \\
      \hline
       Content request arrival rate &  Poisson distribution, $\lambda$ = 10 requests per second \\
      \hline
       Content popularity model & Zipf distribution, $\alpha \in [0.6, 0.8, 1.0, 1.2]$   \\
      \hline
       Routers cache capacity &  [0.05 to 0.25]\% of catalog size  \\
      \hline
       Content placement &  Uniform distribution \\
      \hline
      Replacement strategy & LRU, Random, and PLFU \\
      \hline
       Experiment repetitions & 10 \\
      \hline
       Performance metrics & Cache Hit Ratio, Average Latency, and \\ & Average Hit Distance\\
      \hline
       Benchmark schemes & LCE, DPCP, MNU, PaCC, CRUS, and PePC\\
      \hline
    \end{tabular}}
  \vspace{-2mm}
\end{table}

\subsection{Performance Metrics} In order to evaluate the effectiveness of our proposed strategy \texttt{\name}, we use the most commonly used network performance metrics \cite{Amadeo2021, Huang2019} in NDN/ICN caching domains, which are illustrated below.

\noindent\texttt{1. Cache Hit Ratio:} This metric is used to assess the performance of caching techniques. The cache hit ratio defines the fraction of contents served from the router's cache out of the total number of consumer-generated requests. A higher cache ratio implies that more content is served from the router cache rather than the original content source, which helps to reduce the burden on the original content source as well as the content fetching time. The overall cache hit ratio (denoted as $\overline{\rm CR}$) is computed during the simulation using Equation (\ref{cachehit}). 
\begin{equation} \label{cachehit}
 \overline{\rm CR} = \frac {R_i}{N}
\end{equation}
Where $N$ denotes the total number of consumer-generated requests in the network, and $R_i$ denotes the fraction of content requests served from the router's cache.
\newline\noindent\texttt{2. Average Latency:} This metric measures the total time it takes for a consumer request to reach the content provider (router/original source) and for the content provider to respond to the consumer. The average latency (denoted as $\overline{\rm AL}$) is estimated during the simulation using Equation (\ref{delay}).
\begin{equation} \label{delay}
 \overline{\rm AL} = \frac{ \sum_{i=1}^{N} {D_i}} {N}
\end{equation}
Where $N$ denotes the total number of consumer-generated requests in the network, and $D_i$ denotes the total time required to satisfy $i^{th}$ consumer requests.
\newline\noindent\texttt{3. Average Hit Distance:} This metric measures the average number of hop count Data packets travel from the content provider to the consumer. A lower hit distance indicates more efficient content delivery through nearby routers, resulting in reduced network load. The Average Hit Distance (denoted as $\overline{\rm AD}$) is estimated during the simulation using Equation (\ref{distance}).
\begin{equation} \label{distance}
 \overline{\rm AD} = \frac{ \sum_{i=1}^{N} {H_i}} {N}
\end{equation}
Where $H_i$ represents the number of hops taken by the $i^{th}$ Data packet, and $N$ denotes the total number of Data packets delivered by content providers in response to consumer requests.

\subsection{Network Topology} We evaluate the effectiveness of our proposed \texttt{\name} techniques through simulations on two RocketFuel ISP \cite{Spring2002} topologies: Exodus (AS 3967), comprising 161 nodes, and AboveNet (AS 6461), with 282 nodes. The Icarus simulator uses the Fast Network Simulation Setup (FNSS) \cite{Saino2013a} tool for configuring the network topology. Every topology in this configuration consists of three types of nodes: consumers, routers, and sources. Consumers and sources are artificial nodes with a degree of one, attached to routers with degrees greater than one. Each router is connected to a consumer node, ensuring that every router has a corresponding consumer, while 5\% of the routers with the highest degrees are linked to source nodes. In our simulations, we set a link delay of 0  milliseconds for connections between artificially attached consumers and routers, as well as between source nodes and routers, while the link delays among all the routers in these topologies are derived from the RocketFuel dataset available in the Icarus \cite{Saino2014} simulator. We integrated the Python-based Louvain community recognition technique into Icarus to divide the topology into communities. Equation \ref{clusters} is used to determine the number of communities in each topology. We configured the $\tau$ value to 0.15 for both topologies across all simulation results, except in the case of experiments involving varying community sizes, where we explored different community size configurations. 
\begin{equation} \label{clusters}
 N = \lceil  \tau \cdot V \rceil
\end{equation}
where $N$ is the total number of communities, $V$ is the total number of nodes in the topology, and $\tau$ represents the scaling factor used to determine the number of communities. 

\begin{figure}[!bthp]
\centering
 \vspace{-4mm}
\includegraphics[width= 1.0\columnwidth]{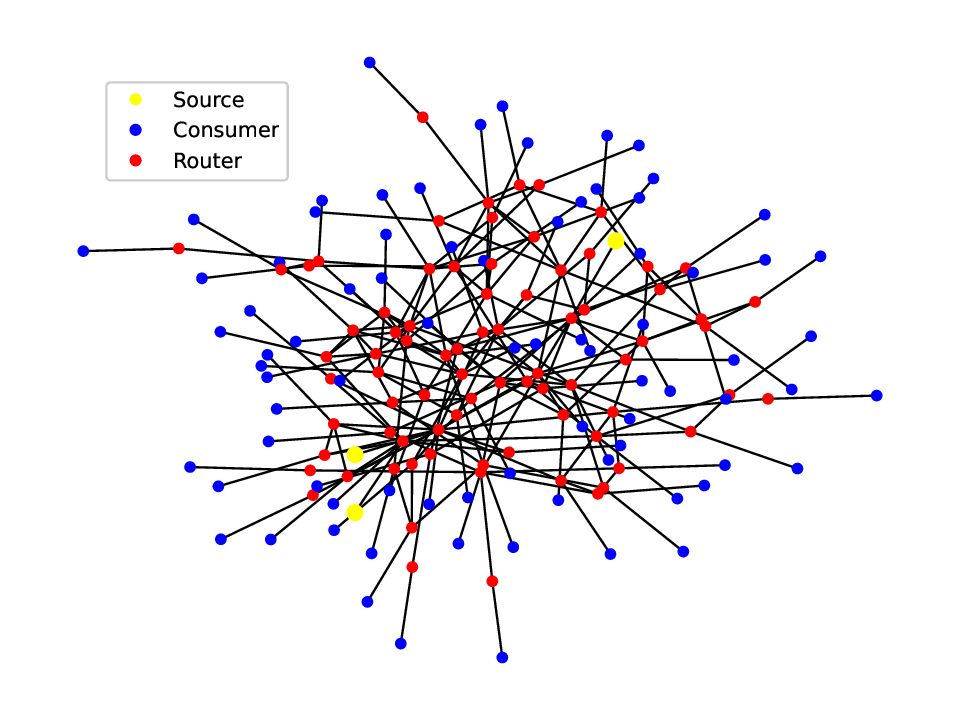}
  \caption{\label{fig:exodusmap} RocketFuel ISP Map Exodus Topology}
  \vspace{-4mm}
\end{figure}

Further details regarding both topologies are provided in Table \ref{tab: topologyinformation}. Fig.~\ref{fig:exodusmap} illustrates the structure of Exodus topology, which includes consumer, source, and router nodes.

\begin{table}[htbp]
\centering
\caption{Network Topologies Details}
\label{tab: topologyinformation}
\scalebox{0.8}{
    \begin{tabular}{|c|c|c|c|c|c|}  
     \hline
      \textbf{Topology} & \textbf{Nodes} & \textbf{Links} & \textbf{Consumers} & \textbf{Sources} & \textbf{Routers} \\
      \hline
      Exodus (AS 3967) & 161 & 229 & 79 & 3 & 79 \\
      \hline
      Abovenet (AS 6461) & 282 & 516 & 138 & 6 & 138 \\
      \hline  
    \end{tabular} }
  \vspace{-4mm}
\end{table}

\subsection{Simulation Results} Here, we present numerical simulation results to evaluate the performance of the proposed \texttt{\name} method. These results consider various factors such as cache sizes, content popularity distribution, community sizes, and the analysis of results with different replacement strategies.

\subsubsection{Impact of the router cache sizes} 
\begin{figure}[htbp]
\begin{subfigure}{0.5\columnwidth}
    \centering
    \includegraphics[width=\textwidth]{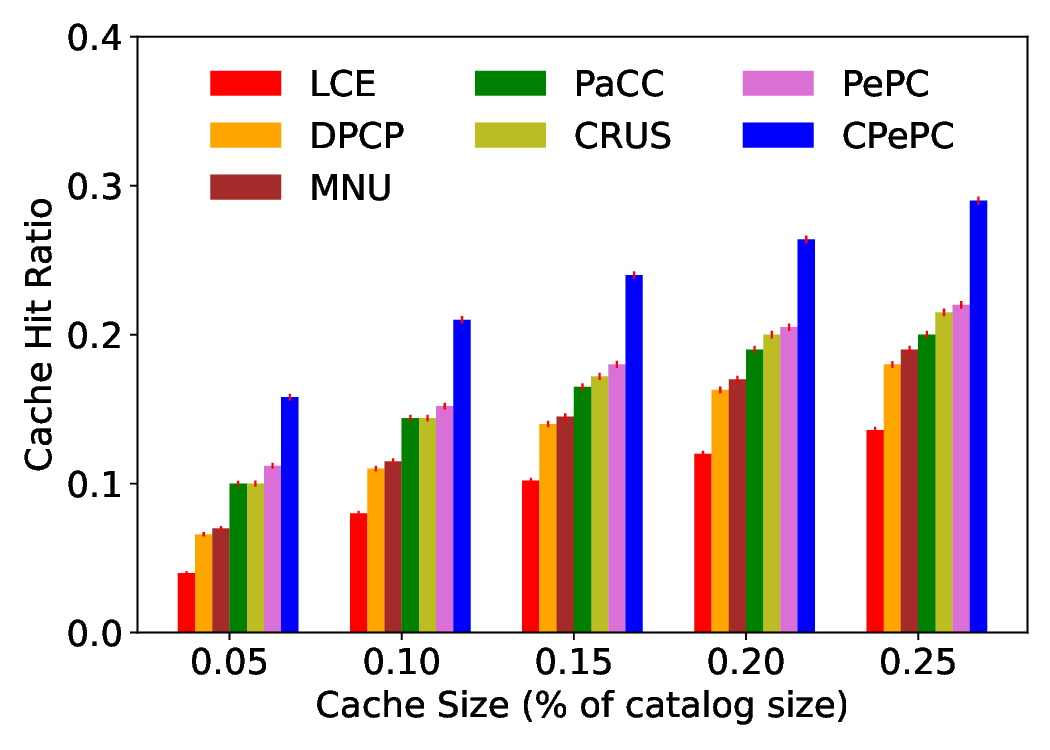}
    \caption{AS 3967}
    \label{fig:cachehitcomparison1}
\end{subfigure}%
\begin{subfigure}{0.5\columnwidth}
    \centering
    \includegraphics[width=\textwidth]{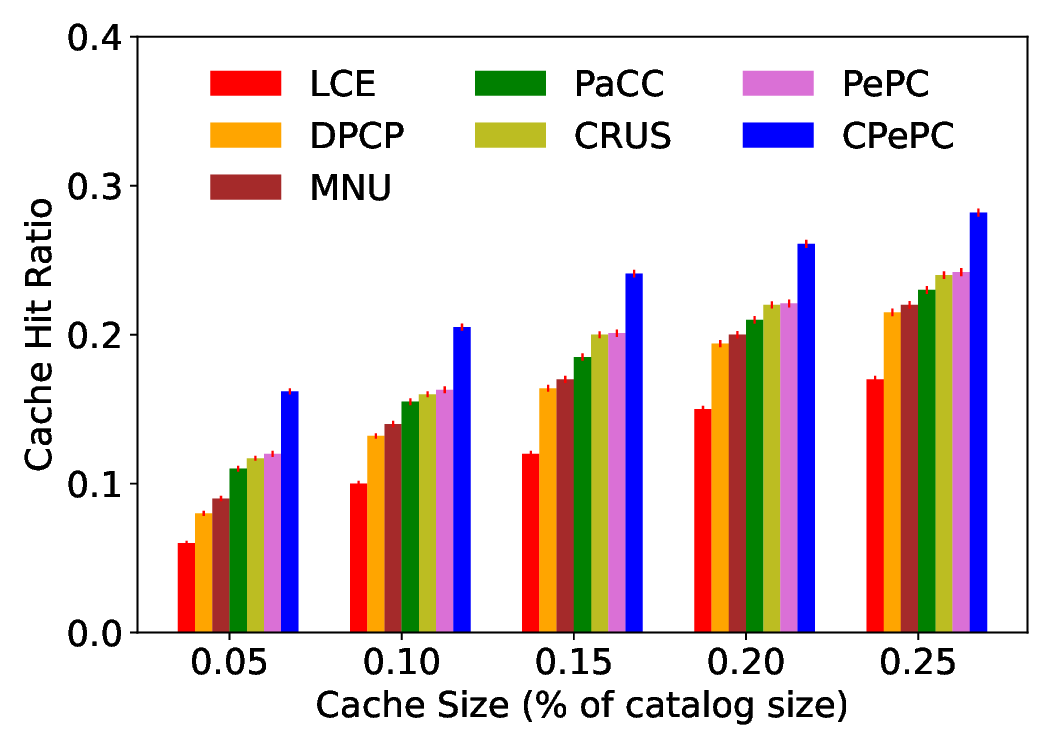}
    \caption{AS 6461}
    \label{fig:cachehitcomparison2}
\end{subfigure}
\noindent \caption{Cache Hit Ratio for Different Cache Sizes on Different Network Topologies ( Zipf $\alpha$= 0.8).}
\label{fig:cachehitcomparison}
\vspace{-6mm}
\end{figure}

Here, we evaluate the performance of all seven caching techniques by varying the cache size from 0.05\% to 0.25\% of the catalog size while keeping the Zipf $\alpha$ value fixed at 0.8. \\
\indent Figs.~\ref{fig:cachehitcomparison1} and \ref{fig:cachehitcomparison2} show that increasing the cache size increases the cache hit ratio of all seven methods on both topologies. A larger cache size enables routers to cache more content. 
When the cache size is relatively small (\textit{i.e.,} 0.05\% of catalog size), we observe cache hit ratio improvements of up to 41\% on AS 3967 and up to 35\% on AS 6461, compared to PePC (second-best method) and up to 58\% on AS 3967 and up to 39\% on AS 6461, compared to CRUS. Further, even with larger cache sizes (\textit{i.e.,} 0.25\% of catalog size), we observe improvements up to 31.8\% on AS 3967 and 16.5\% on AS 6461 compared to PePC, 34.9\% on AS 3967, and 17.5\% on AS 6461 as compared to CRUS. The greater improvement in hit rates is noticeable with smaller cache sizes, as opposed to larger cache sizes, because as cache sizes increase, all strategies tend to accommodate more content in network routers. 

\begin{figure}[htbp]
\begin{subfigure}{0.5\columnwidth}
    \centering
    \includegraphics[width=\textwidth]{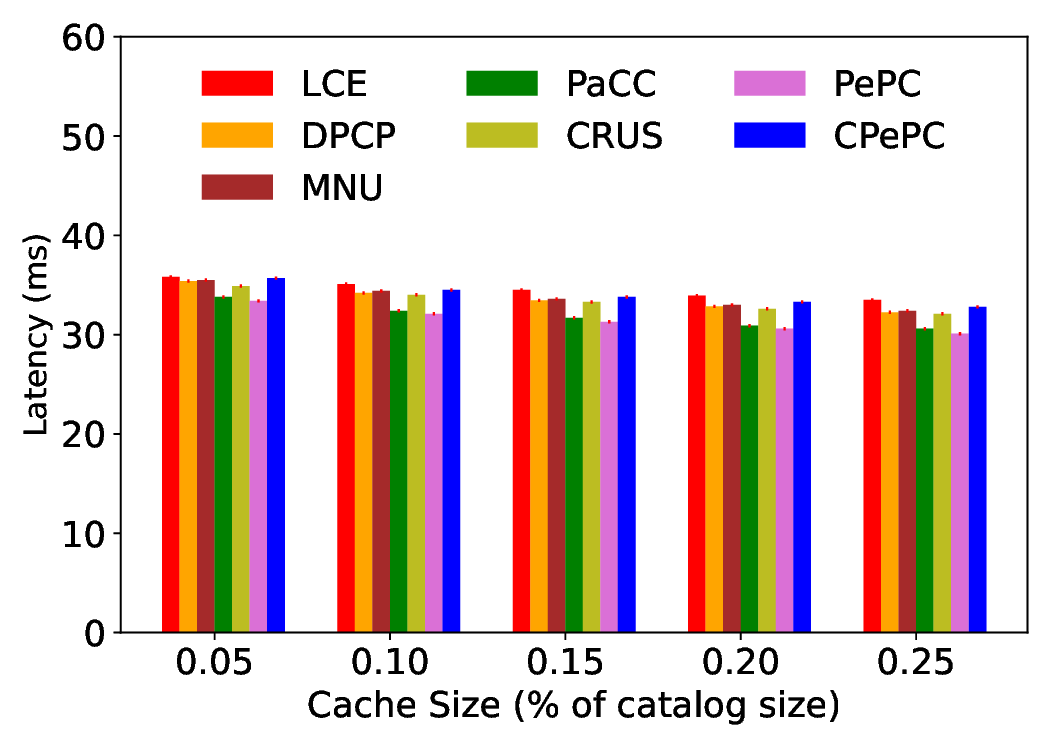}
    \caption{AS 3967}
    \label{fig:latencycomparison1}
\end{subfigure}%
\begin{subfigure}{0.5\columnwidth}
    \centering
    \includegraphics[width=\textwidth]{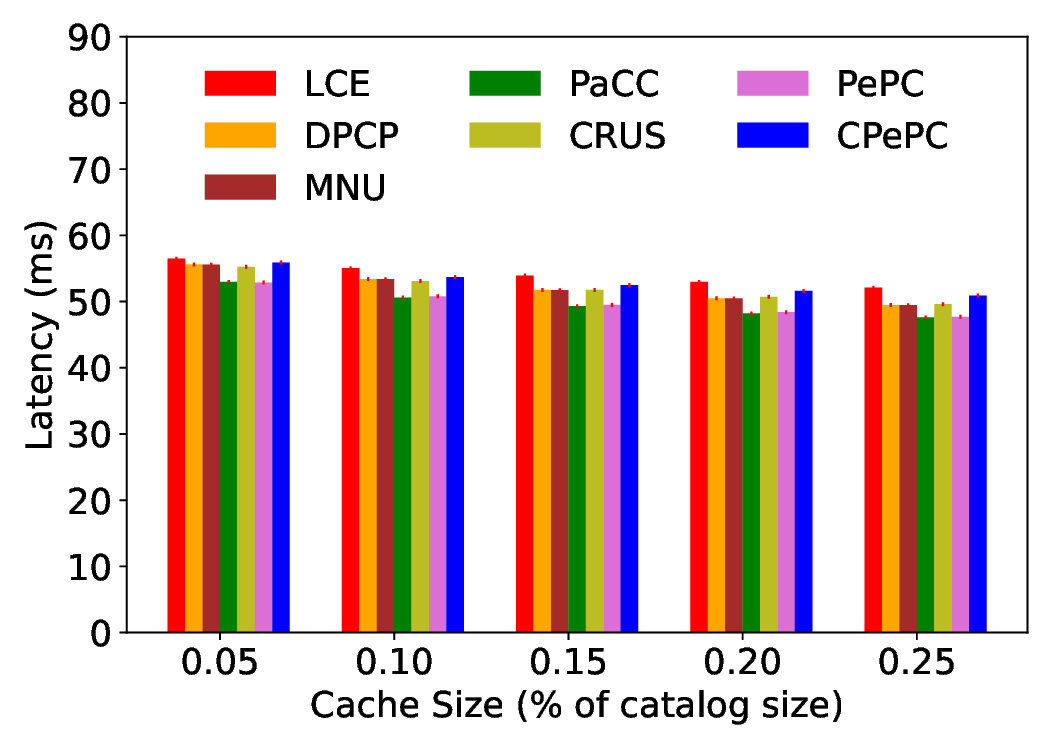}
    \caption{AS 6461}
    \label{fig:latencycomparison2}
\end{subfigure}
\noindent \caption{Average Latency (in milliseconds) for Different Cache Sizes on Different Network Topologies (Zipf $\alpha$= 0.8).}
\label{fig:latencycompariosn}
\vspace{-2mm}
\end{figure}

\indent Fig.~\ref{fig:latencycompariosn} illustrates the impact of latency while increasing the cache size for all seven methods across both topologies. As the cache size increases, the average latency for all seven methods decreases because content is served from nearby routers. Compared to the second-best method, PePC, the average latency of the proposed \texttt{\name} increases by up to 8.2\% on AS 3967 and up to 6.3\% on AS 6461. However, the average latency of the \texttt{\name} strategy is up to 2.1\% lower on AS 3967 and up to 2.6\% lower on AS 6461 compared to LCE. The higher content access time in the \texttt{CPePC} method, compared to the PePC, CRUS, and PaCC on-path caching methods, is due to \texttt{CPePC} relying on the leader node to search for content within the community, which introduces delays in content retrieval.

\begin{figure}[htbp]
\begin{subfigure}{0.5\columnwidth}
    \centering
    \includegraphics[width=\textwidth]{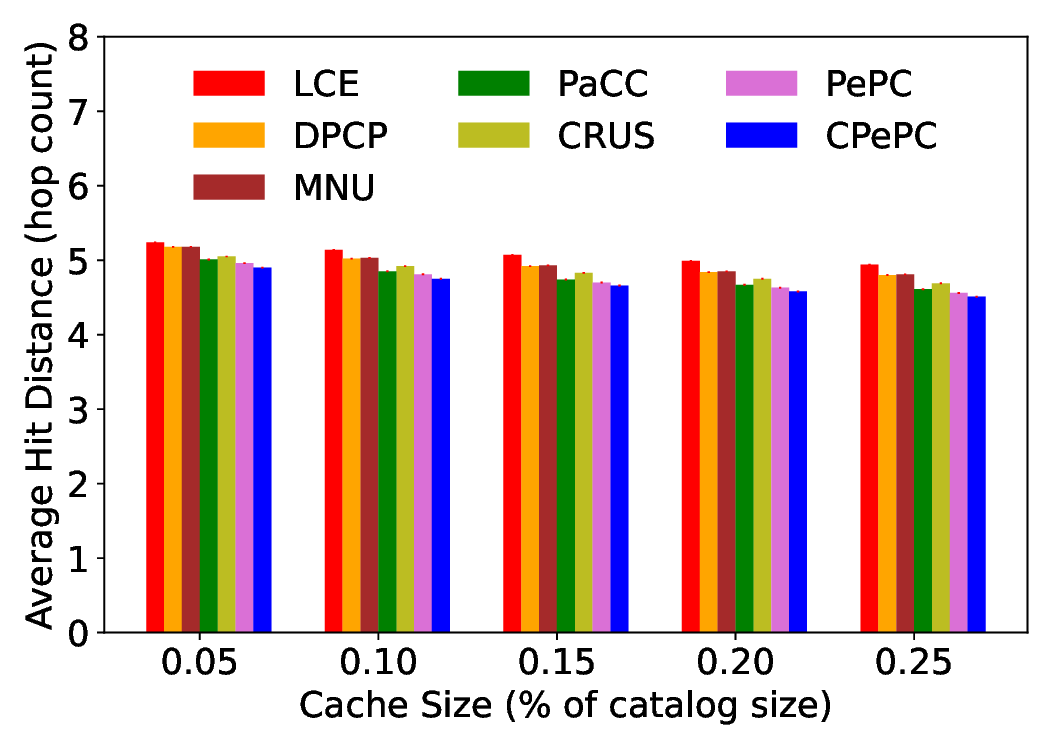}
    \caption{AS 3967}
    \label{fig:hopcountcomparison1}
\end{subfigure}%
\begin{subfigure}{0.5\columnwidth}
    \centering
    \includegraphics[width=\textwidth]{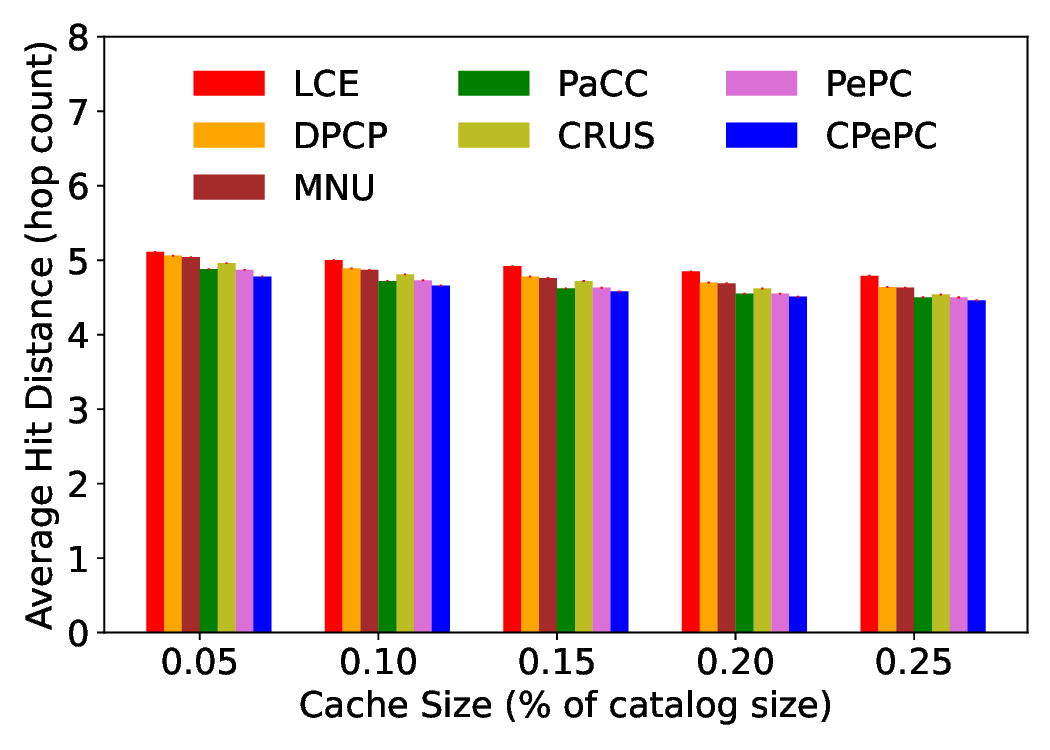}
    \caption{AS 6461}
    \label{fig:hopcountcomparison2}
\end{subfigure}
\noindent \caption{Average Hit Distance for Different Cache Sizes on Different Network Topologies ( Zipf $\alpha$= 0.8).}
\label{fig:hopcountcomparison}
\vspace{-4mm}
\end{figure}

\indent Fig.~\ref{fig:hopcountcomparison} illustrates the relationship between caching strategies and average hit distance in terms of hop count while varying cache sizes. A lower hit distance means that the Data packet travels fewer hops between the provider and the consumer. As cache size increases, the hit distance decreases for all caching methods because content is more readily available in nearby routers. Notably, our proposed \texttt{CPePC} strategy consistently outperforms the other six caching techniques across all cache sizes. On the AS 3967 topology, \texttt{CPePC} achieves a hop count reduction of up to 8.7\% compared to LCE, 3.8\% compared to CRUS, 2.2\% compared to PaCC, and 1.27\% compared to PePC. Similarly, on the AS 6461 topology, \texttt{CPePC} outperforms other strategies with reductions of up to 6.9\% compared to LCE, 3.6\% compared to CRUS, 2.1\% compared to PaCC, and 1.87\% compared to PePC. This superior performance of the \texttt{CPePC} method can be attributed to better cooperative routing and caching within the community.

\subsubsection{Impact of the content popularity} 
\begin{figure}[htbp]
\begin{subfigure}{0.5\columnwidth}
    \centering
    \includegraphics[width=\textwidth]{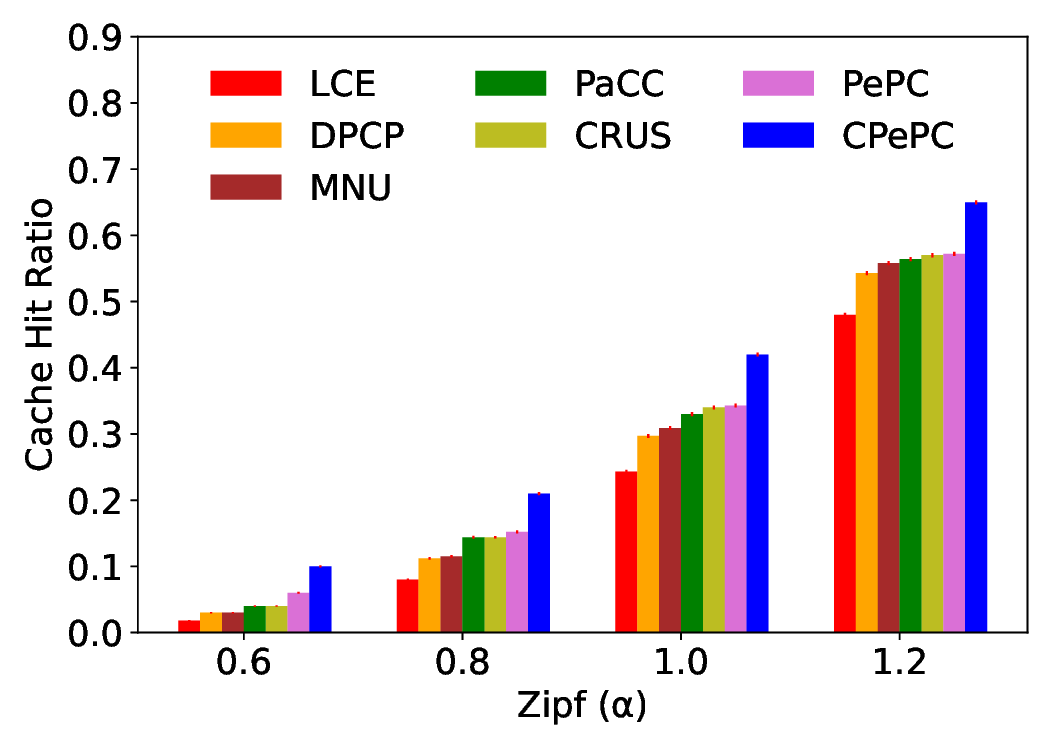}
    \caption{AS 3967}
    \label{fig:alphacomparison1}
\end{subfigure}%
\begin{subfigure}{0.5\columnwidth}
    \centering
    \includegraphics[width=\textwidth]{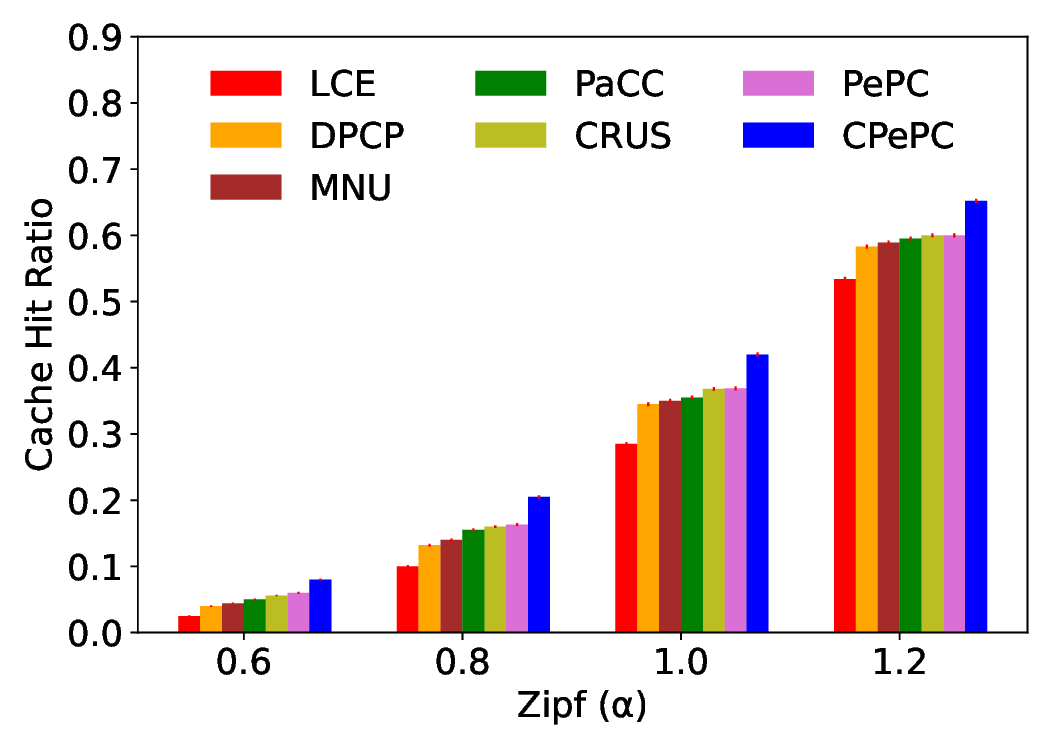}
    \caption{AS 6461}
    \label{fig:alphacomparison2}
\end{subfigure}
\noindent \caption{Cache Hit Ratio with Different Zipf $\alpha$ on Different Network Topologies (Cache Size= 0.1\%).}
\label{fig:alphacompariosn}
\vspace{-4mm}
\end{figure}

The popularity of content might change dramatically over time. To understand how different caching techniques perform under different popularity conditions, we vary the Zipf $\alpha$ value from 0.6 to 1.2, which controls the content request patterns. A smaller $\alpha$ corresponds to a situation where a large number of contents have similar request frequencies, while higher $\alpha$ values indicate that a small subset of contents receives a significant number of requests, while the majority receive relatively few. \\ \indent Fig.~\ref{fig:alphacompariosn} illustrates that as the $\alpha$ value increases, the cache hit ratio of all seven caching techniques increases significantly. This is attributed to a few contents that heavily dominate the request patterns, and these frequently requested contents are present in the router. Notably, our \texttt{\name} consistently achieves a higher cache hit ratio on both topologies, regardless of content popularity, when compared to the other six methods. Figs.~\ref{fig:alphacomparison1} and \ref{fig:alphacomparison2} show this improvement, with an increase in cache hit ratio of up to 66.6\% on AS 3967 and up to 33.3\% on AS 6461 as compared to the PePC strategy. 

\subsubsection{Impact of the catalog sizes} Here, we present the simulation results illustrating the impact of catalog size (number of distinct contents) on the cache hit ratio. For this evaluation, each router is allocated a fixed cache size of 10 slots, indicating that each router can store a maximum of 10 contents in its cache, and Zipf $\alpha$ is set to 0.8.

\begin{figure}[htbp]
\begin{subfigure}{0.5\columnwidth}
    \centering
    \includegraphics[width=\textwidth]{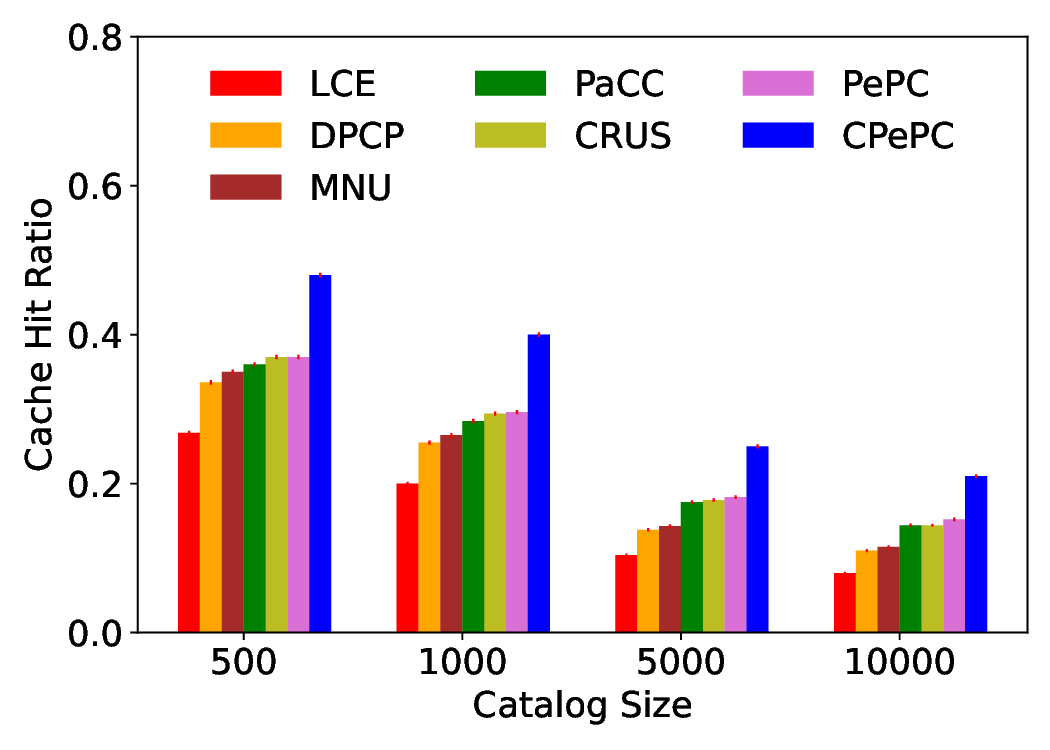}
    \caption{AS 3967}
    \label{fig:catalogcomparison1}
\end{subfigure}%
\begin{subfigure}{0.5\columnwidth}
    \centering
    \includegraphics[width=\textwidth]{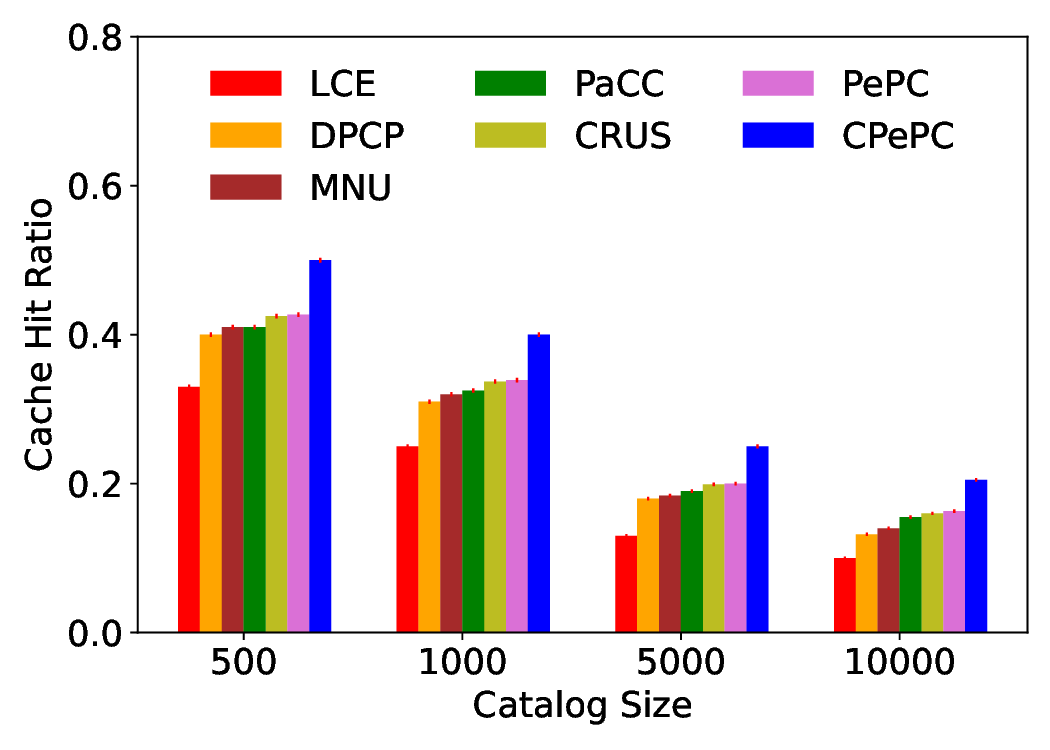}
    \caption{AS 6461}
    \label{fig:catalogcomparison2}
\end{subfigure}
\noindent \caption{Cache Hit Ratio with Varying Catalog Sizes Across Different Network Topologies (Zipf $\alpha$=0.8).}
\label{fig:catalogcompariosn}
\vspace{-4mm}
\end{figure}

\indent Fig.~\ref{fig:catalogcompariosn} depicts the effect of catalog size on the cache hit ratio for all seven strategies across both topologies. Figs.~\ref{fig:catalogcomparison1} and \ref{fig:catalogcomparison2} show that when the catalog size is small, the cache hit ratio increases for all the caching techniques. However, the cache hit ratio declines when the catalog size is large. This is due to the fact that with a larger catalog size, consumers request a more diverse range of content, making it less likely for all content to be available in the routers due to their limited capacity. In contrast, with a smaller catalog size, a higher proportion of consumer requests can be satisfied from the routers. Notably, our proposed \texttt{\name} technique outperformed the other six caching methods regardless of catalog size. The higher cache hit ratio of \texttt{\name} is due to the cooperation among routers. \texttt{\name} eliminates content duplication within the community, improving network-wide diversity regardless of the number of contents.   

\subsubsection{Impact of the community sizes} 
\begin{figure}[htbp]
\begin{subfigure}{0.5\columnwidth}
    \centering
    \includegraphics[width=\textwidth]{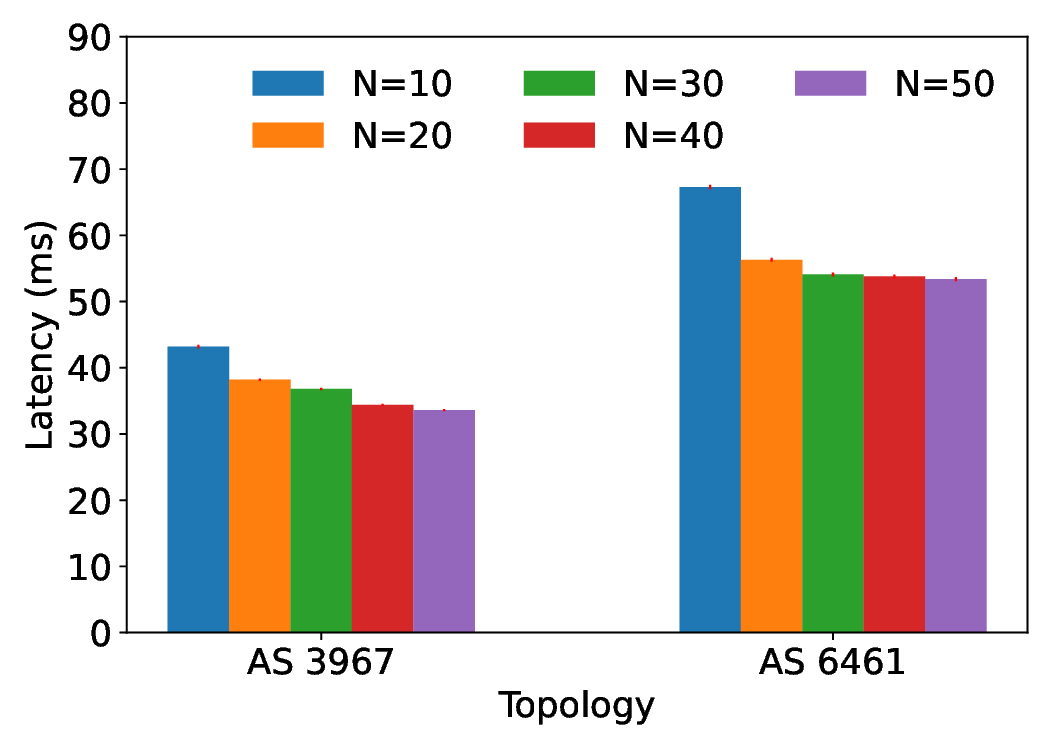}
    \caption{Average Latency}
    \label{fig:communitylatencycomparison}
\end{subfigure}%
\begin{subfigure}{0.5\columnwidth}
    \centering
    \includegraphics[width=\textwidth]{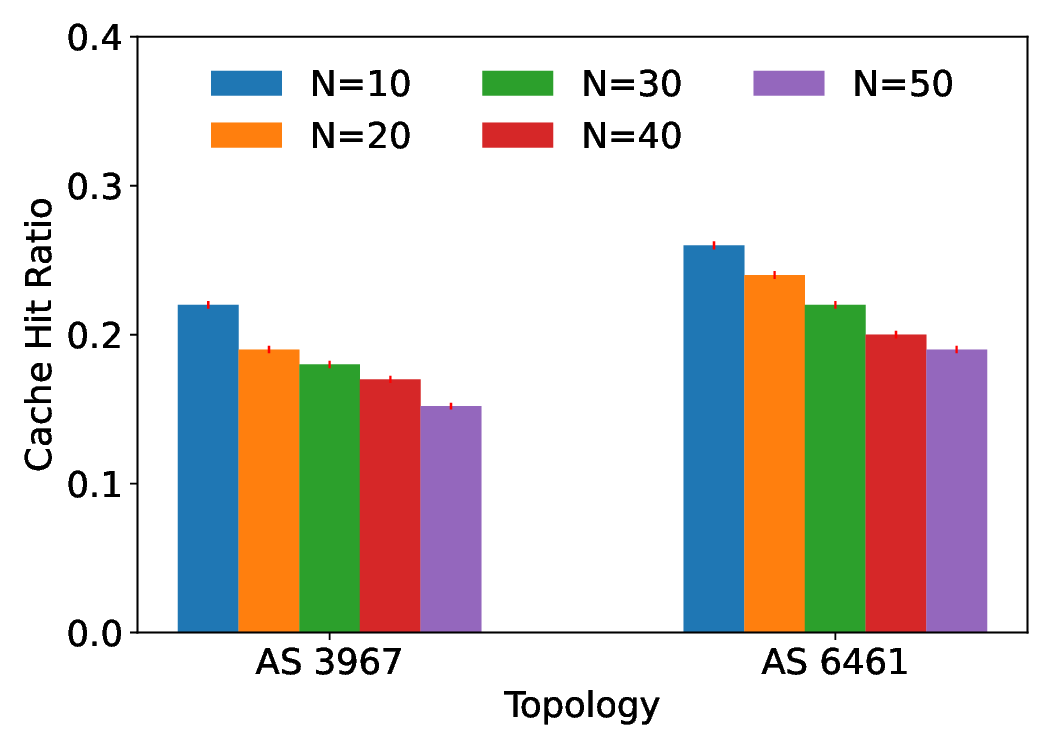}
    \caption{Cache Hit Ratio}
    \label{fig:communitycachecomparison}
\end{subfigure}
\noindent \caption{\texttt{\name} Performance on Different Topologies with Varying Community Sizes (\texttt{N=10} to \texttt{N=50}) with fixed Cache Size=0.1\% and Zipf $\alpha$=0.8.}
\label{fig:communitycompariosn}
\vspace{-4mm}
\end{figure}

Here we study the impact of community size on the performance of \texttt{\name}. Fig.~\ref{fig:communitycompariosn} shows the impact of varying community sizes on both topologies with respect to the average latency and cache hit ratio. We vary the number of communities from $N=10$ to $50$ while keeping fixed cache size=0.1\% and $\alpha$=0.8.  The number of communities can be varied by setting different values to resolution parameter in the Louvain algorithm \cite{Blondel2008}. When the resolution value is below 1, the algorithm generates larger-sized communities, whereas if it exceeds 1, it generates smaller-sized communities. A large number of communities implies fewer nodes in each community. 

\indent Fig.~\ref{fig:communitylatencycomparison} depicts the average latency of the \texttt{\name} across both topologies. From Fig.~\ref{fig:communitylatencycomparison}, we can see that latency decreases as the number of communities increases. This trend is due to the fact that an increase in the number of communities reduces the size of individual communities, bringing the leader closer to other routers in the community, and thereby reducing content search delays within the community. On the other hand, fewer communities result in larger community sizes, leading to higher latency when searching for content through the leader. From Fig.~\ref{fig:communitycachecomparison}, we can observe that the \texttt{\name} achieves a higher cache hit ratio when the number of communities is less. However, as the number of communities increases, the cache hit ratio decreases significantly. The decrease in cache hit ratio with the increasing number of communities is due to the same content being cached at multiple communities. This occurs because the \texttt{\name} method ensures zero redundancy within each community, but the same content may be cached across different communities. Hence, when content is transmitted between content providers and consumers, it may traverse a larger number of communities, resulting in content redundancy due to storing the same content in multiple communities and subsequently reducing the cache hit ratio. Choosing an appropriate community size is thus crucial for achieving a balance between cache hit ratio and latency. 

\subsubsection{Impact of the replacement policies} 
\begin{figure}[htbp]
\begin{subfigure}{0.5\columnwidth}
    \centering
    \includegraphics[width=\textwidth]{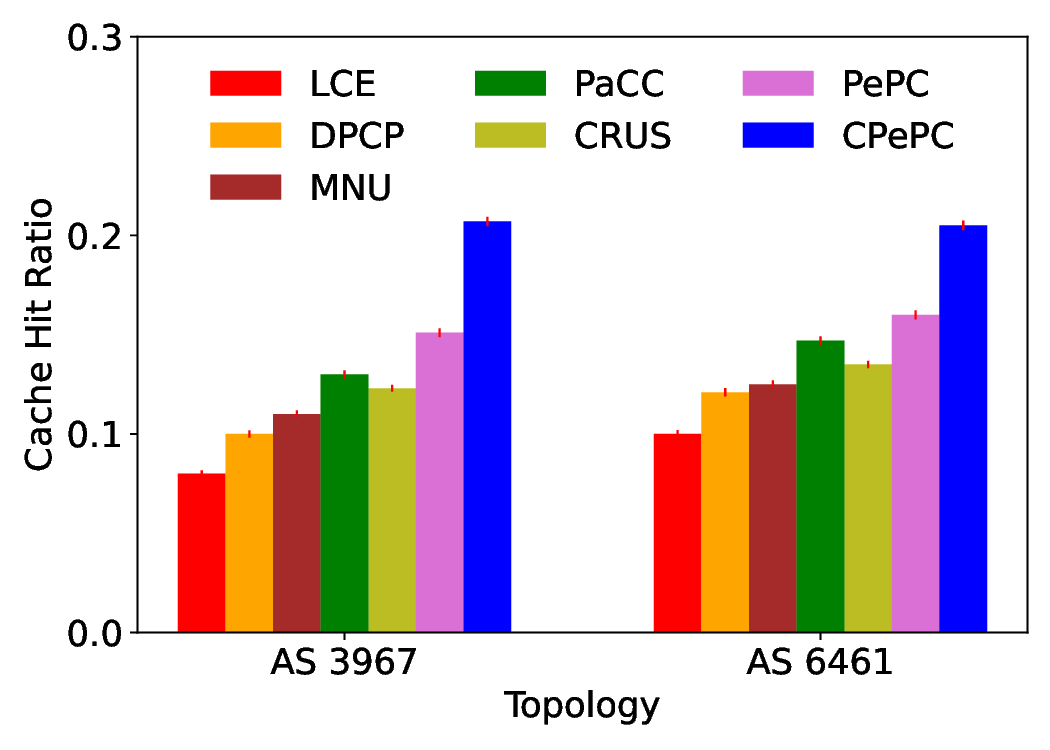}
    \caption{Random}
    \label{fig:randomcomparison1}
\end{subfigure}%
\begin{subfigure}{0.5\columnwidth}
    \centering
    \includegraphics[width=\textwidth]{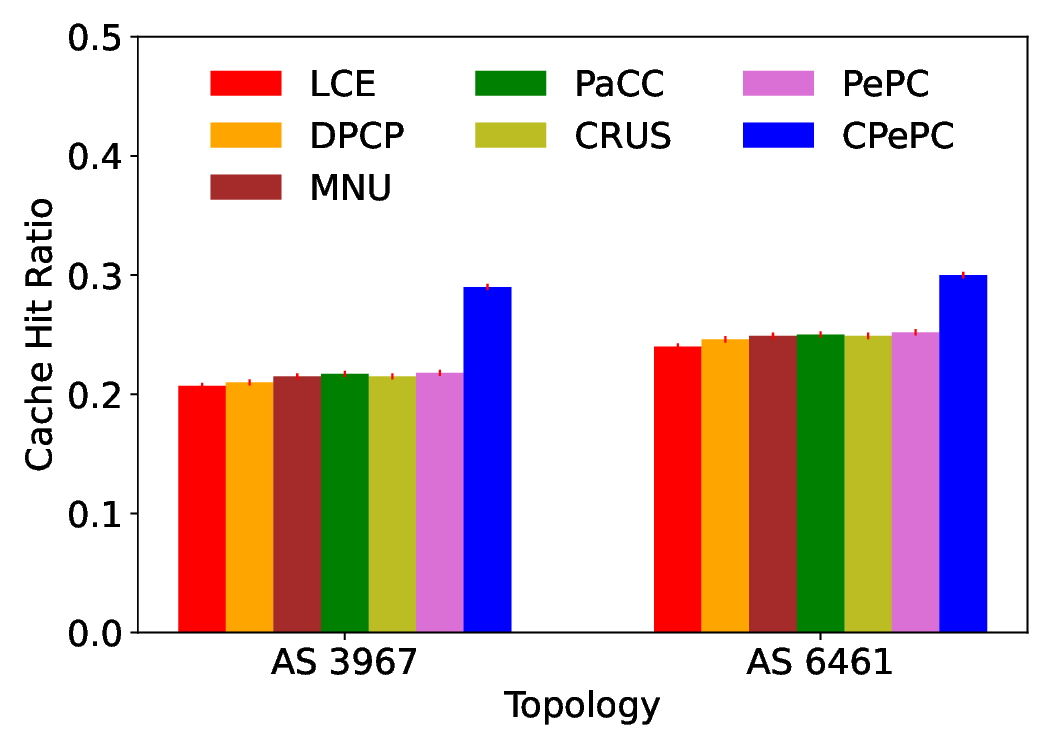}
    \caption{PLFU}
    \label{fig:plfucomparison2}
\end{subfigure}
\noindent \caption{Cache Hit Ratio Comparison with Different Replacement Policies on Different Network Topologies (cache size=0.1\%, Zipf $\alpha$=0.8).}
\label{fig:replacementcompariosn}
\vspace{-6mm}
\end{figure}

So far, all of the aforementioned simulation results have been reported using the LRU policy. In this specific part, we extend our analysis by comparing all seven caching methods using various replacement policies on both network topologies. This comparison allows us to evaluate how our proposed \texttt{\name} technique performs in contrast to the other six caching techniques under different replacement policies. \\ 
\indent Fig.~\ref{fig:replacementcompariosn} depicts the cache hit ratio of different caching methods employing two different replacement policies. Fig.~\ref{fig:randomcomparison1} shows the performance of caching techniques using the Random replacement strategy, in which content is randomly selected from the router to make way for new content when router capacity is full. Fig.~\ref{fig:plfucomparison2} shows the performance analysis using the Perfect-LFU (PLFU), in which content is evicted based on frequency, with lower-frequency content being removed from the router upon the arrival of new content. Fig.~\ref{fig:replacementcompariosn} demonstrates that the proposed \texttt{\name} method outperforms other caching techniques on both topologies, regardless of the replacement policies considered for content eviction.

\subsubsection{Impact of the caching threshold parameters} 
\begin{figure}[htbp]
\begin{subfigure}{0.5\columnwidth}
    \centering
    \includegraphics[width=\textwidth]{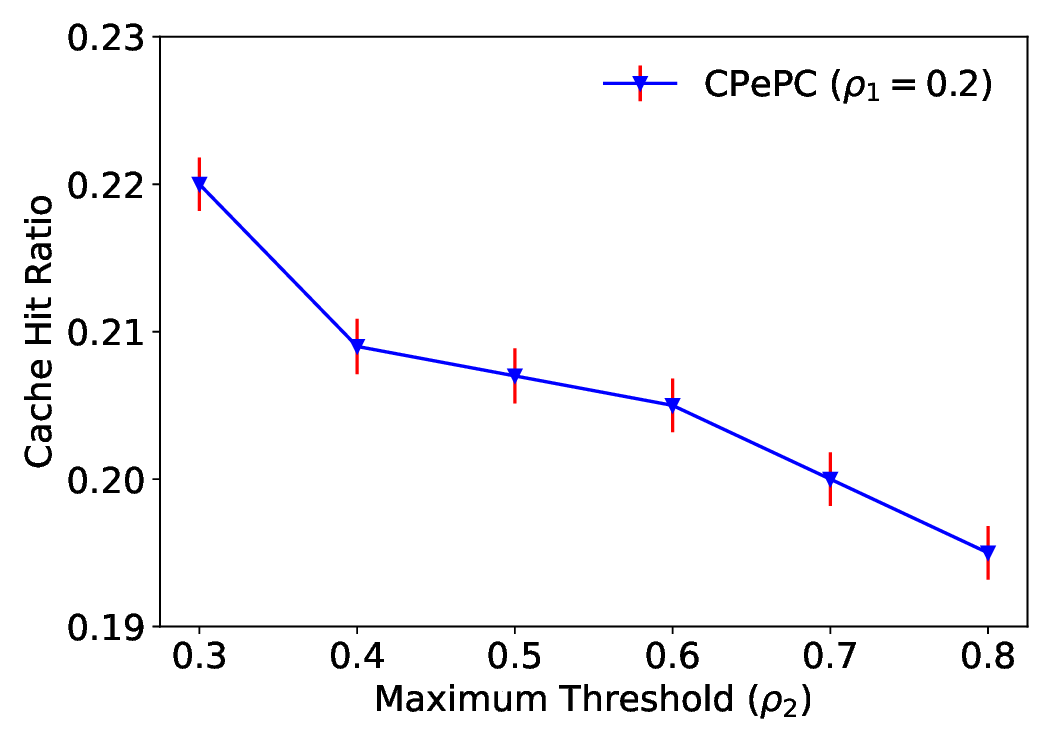}
    \caption{Cache Hit Ratio}
    \label{fig:maxvarycomparison1}
\end{subfigure}%
\begin{subfigure}{0.5\columnwidth}
    \centering
    \includegraphics[width=\textwidth]{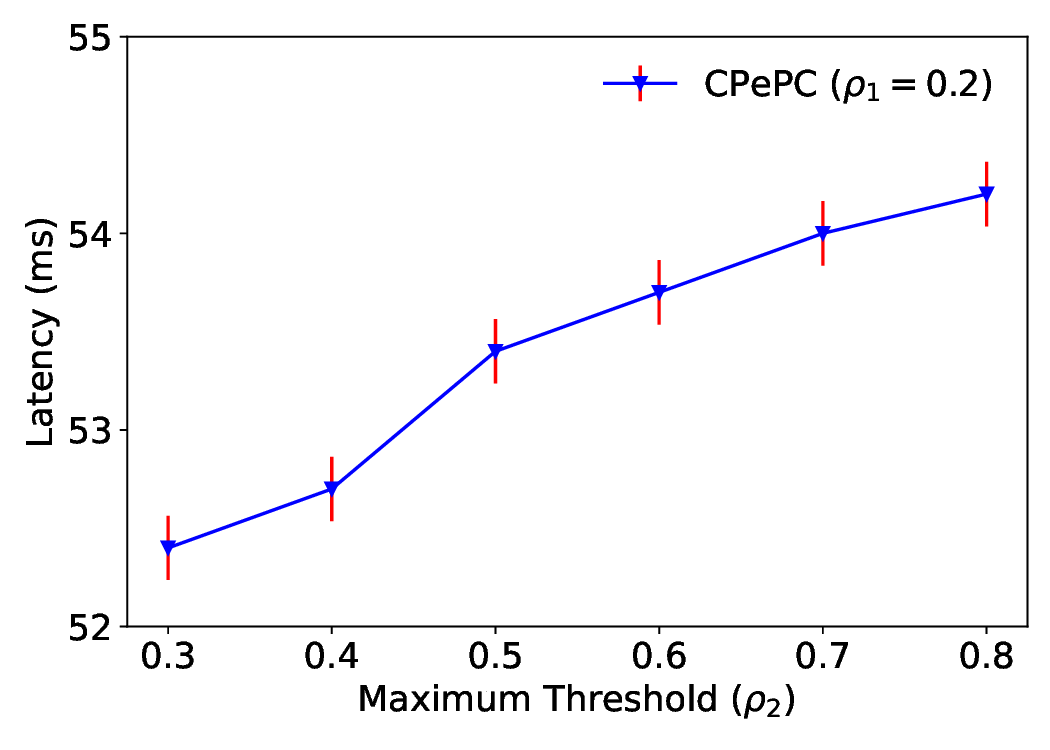}
    \caption{Average Latency}
    \label{fig:maxvarycomparison2}
\end{subfigure}
\noindent \caption{Performance Analysis of Cache Hit Ratio and Average Latency with Varying $\rho_2$.}
\label{fig:maxvarycompariosn}
\vspace{-4mm}
\end{figure}

Here, we explore the impact of adjusting the minimum and maximum caching threshold parameters ($\rho_1$ and $\rho_2$) for \texttt{\name} method in terms of cache hit ratio and average latency while maintaining the fixed $\alpha$ value of 0.8 and cache size of 0.1\% on the AS 6461 topology. Fine-tuning the values of $\rho_1$ and $\rho_2$ plays a crucial role in balancing cache hit ratio and latency. The choice of these values significantly impacts the performance of the caching system.  \\
\indent Fig.~\ref{fig:maxvarycompariosn} illustrates the cache hit ratio and average latency. This analysis involves varying the parameter $\rho_2$ within the range of 0.3 to 0.8 while keeping $\rho_1$ fixed at a value of 0.2. Figs.~\ref{fig:maxvarycomparison1} and \ref{fig:maxvarycomparison2} show that as the $\rho_2$ value increases, the cache hit ratio starts to decrease, and average latency increases. This behavior can be attributed to the fact that with a lower $\rho_2$ value, the average cache occupancy quickly approaches its maximum threshold ($max_{th}$). Thus, content is only cached in the router when its popularity exceeds the dynamic threshold ($\Delta$). This leads to a reduction in the caching of less popular content. 

\begin{figure}[htbp]
\begin{subfigure}{0.5\columnwidth}
    \centering
    \includegraphics[width=\textwidth]{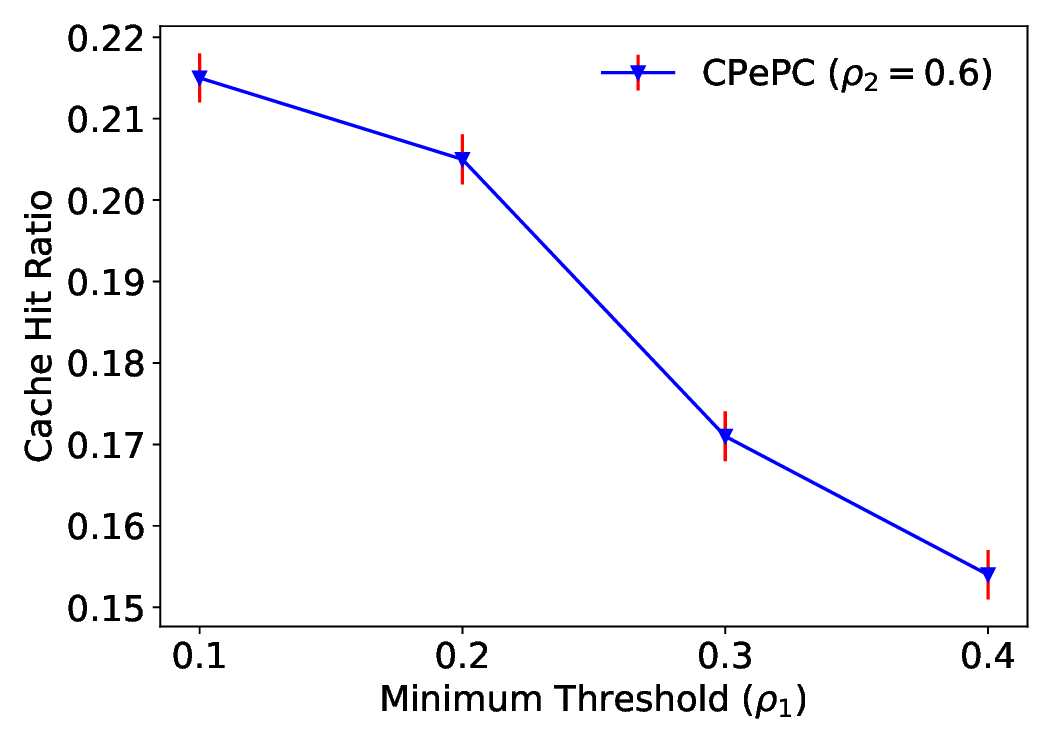}
    \caption{Cache Hit Ratio}
    \label{fig:minvarycomparison1}
\end{subfigure}%
\begin{subfigure}{0.5\columnwidth}
    \centering
    \includegraphics[width=\textwidth]{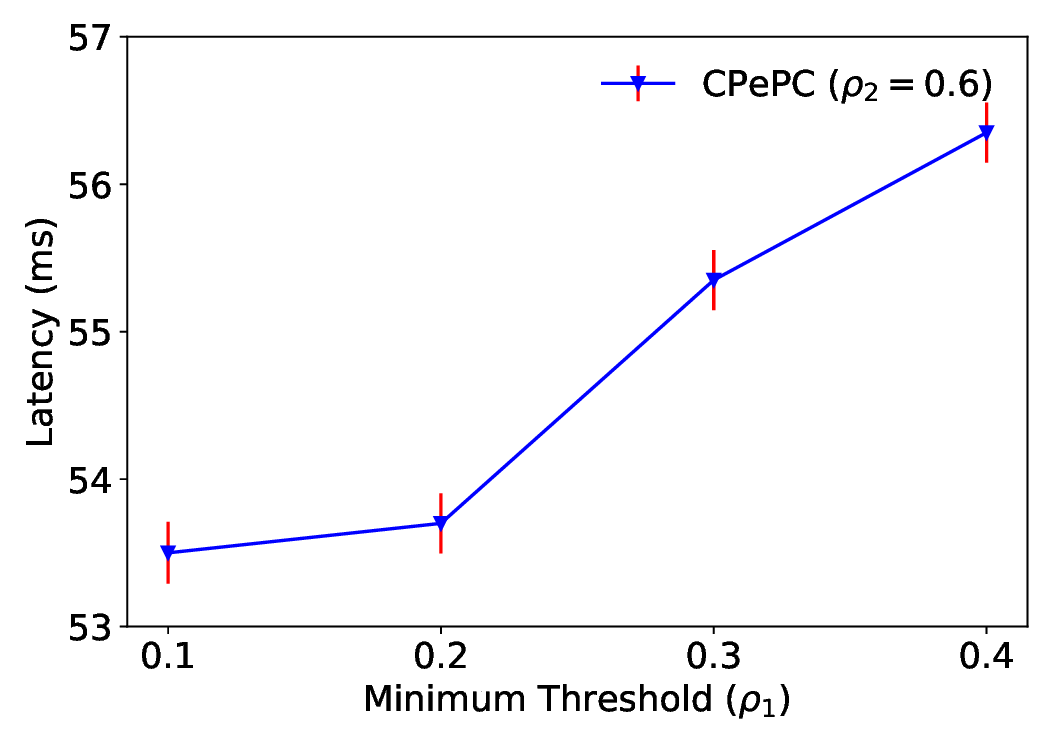}
    \caption{Average Latency}
    \label{fig:minvarycomparison2}
\end{subfigure}
\noindent \caption{Performance Analysis of Cache Hit Ratio and Average Latency with Varying $\rho_1$.}
\label{fig:minvarycompariosn}
\vspace{-4mm}
\end{figure}

Fig.~\ref{fig:minvarycompariosn} illustrates the cache hit ratio and average latency metrics, with $\rho_1$ varying between 0.1 and 0.4 while $\rho_2$ is held constant at 0.6. Figs.~\ref{fig:minvarycomparison1} and \ref{fig:minvarycomparison2} show that as the value of $\rho_1$ increases, the cache hit ratio starts to decrease while average latency increases. This trend is due to the fact that when $\rho_1$ is set to a lower value, the average cache occupancy quickly reaches its minimum threshold occupancy ($min_{th}$). This results in better cache utilization, where initially all content is cached, and later, high-popularity content is stored in the router, potentially displacing less popular content.

\subsection{The Overhead of \texttt{\name} and Deployment Considerations} 

\texttt{\name} performs routing and caching operations by coordinating with the leader node of the respective community, which introduces communication overhead. Here, we look at the overhead associated with implementing the \texttt{\name} technique.

\begin{figure}[htbp]
 \centering
  \includegraphics[width= 1.0\columnwidth]{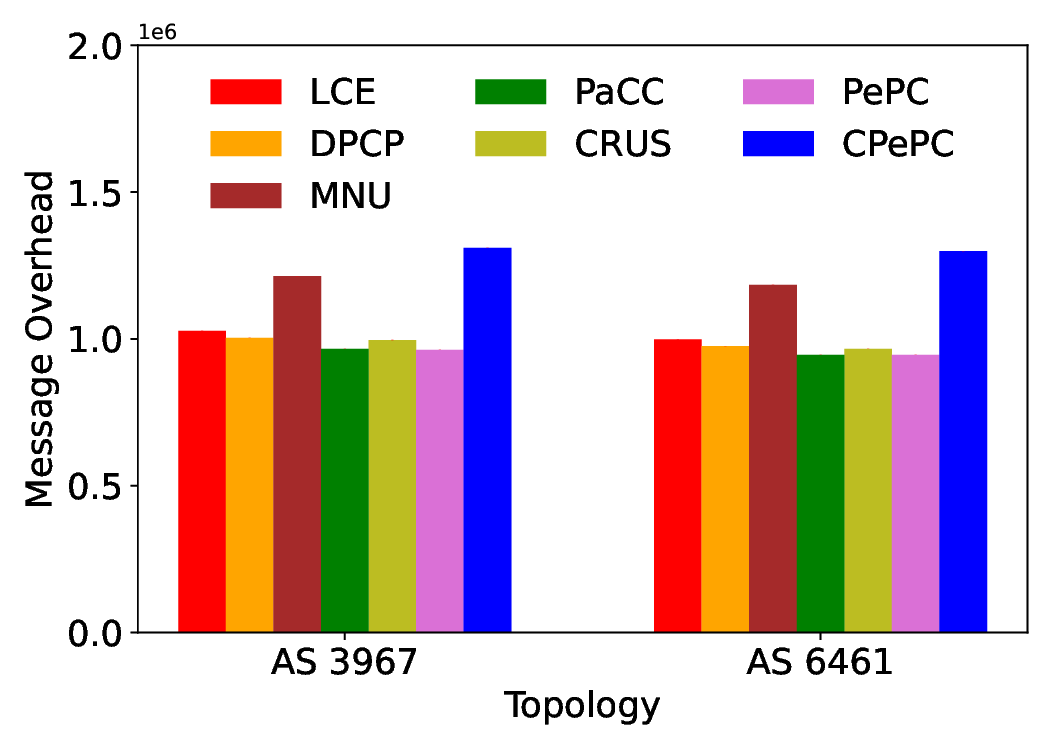}
  \caption{Comparison of Total Number of Messages Exchanged Across Different Network Topologies (cache size=0.1\%, Zipf $\alpha$=0.8).}
   \label{fig:overhead}
   \vspace{-4mm}
\end{figure}

Fig.~\ref{fig:overhead} presents a comparison of the number of messages involved in routing and caching content for all seven caching techniques. The number of messages represents the total data exchanged in the network to retrieve content requested by consumers from the provider. The simulation was conducted on the AS 3967 and AS 6461  topologies, configuring the cache size to 0.1\% and Zipf $\alpha$ to 0.8 while keeping the remaining parameters consistent with those outlined in the previous subsection (\textsl{Simulation Setup}). MNU requires additional messages compared to other on-path techniques (LCE, DPCP, PaCC, CRUS, and PePC) because, in MNU, the content provider coordinates with the network manager to determine the best routers for caching along the content delivery path. On the other hand, the \texttt{\name} includes additional messages that are required for coordination with the community leader nodes. These additional messages include content searching within the community through the leader node while forwarding Interest packets, coordinating with the leader node during content caching, and leader nodes exchanging popularity information with other leader nodes of the communities in the network.

The community division approach of \texttt{\name} improves content distribution and reduces the cache redundancy problem associated with on-path caching strategies. \texttt{\name} also addresses the challenges of operating the network using a centralized controller \cite{Zhang2020} by dividing the large network into smaller groups and selecting a leader within each group. However, these advantages come with increased costs of content searching and caching.

\noindent\textsl{Design complexity:} \texttt{\name} leverages the NDN packet format specification \cite{ndninterestpacket} to add a few additional fields for efficiently searching and caching content within the community. Using the changed packet format will reduce the overhead and communication latency between routers and the community leader. However, the additional fields in the packet structure make lookup and update operations more complex to handle at the router.

\noindent\textsl{Scalability:} The community division approach of \texttt{\name} alleviates the load on a single centralized node (controller) for managing the network and also reduces the burden on other routers for routing and caching decisions. However, having a single leader within each community can lead to potential single points of failure. This issue can be addressed by extending the approach and having backup leader nodes within the community. Another issue that warrants deployment considerations is that of managing the community structure and updating it on a regular basis.

\section{Conclusion} \label{sec5}
In-network caching of content improves the content delivery performance in named data networks.  However, the limited capacity of network routers poses a significant challenge in efficiently caching content to meet the varied demands of users. In this paper, we introduce a community-based predictive cooperative caching scheme named \texttt{\name} for NDN in-network caching to enhance the overall performance. \texttt{\name} divides the network into distinct communities and designates a leader node in each community to facilitate cooperation among the routers. These leader nodes monitor both local and global content popularity, direct consumer requests to potential providers, and ensure no redundancy within their community. \texttt{\name} makes predictive caching decisions based on the current cache occupancy of the router within the community. We compared the caching performance of \texttt{\name} with state-of-the-art caching methods, such as LCE, DPCP, MNU, PaCC, and CRUS, through simulation-based experiments on standard RocketFuel topologies using a discrete event simulator. The comparison is reported in terms of cache hit ratio, average latency, and average hit distance. Our extensive simulation results demonstrate that the \texttt{\name} significantly improves the cache hit ratio and average hit distance compared to state-of-the-art methods, with a little trade-off in access latency. Future work will focus on exploring efficient data structures to share cache information with the leader node and on reducing the time required for content searching.

\bibliographystyle{ieee}
\bibliography{Bibliography}

\end{document}